\documentclass[11pt,twoside]{article}
\pdfoutput=1

\usepackage{amsfonts}
\usepackage{amsmath}
\usepackage{amsthm}
\usepackage{amssymb}
\usepackage{amscd}
\usepackage{cite}
\usepackage{eucal}
\usepackage{bm}
\usepackage{graphicx}
\usepackage{a4}
\usepackage{microtype}
\usepackage{times}
\usepackage{float}

\input{macro.lib}

\renewcommand{\appendix}{%
   \renewcommand{\section}{
        \secdef\Appendix\sAppendix}%
   \setcounter{section}{0}%
   \renewcommand{\thesection}{\Alph{section}}%
   \renewcommand{\theequation}{\thesection.\arabic{equation}}%
}

\newcommand{\Appendix}[2][?]{%
     \refstepcounter{section}%
     \setcounter{equation}{0}%
     \addcontentsline{toc}{appendix}%
          {\protect\numberline{\appendixname~\thesection} #1}%
     \vspace{\baselineskip}%
     {\noindent\large\bfseries\appendixname\ \thesection: #2\par}%
     \sectionmark{#1}\vspace{\baselineskip}}

\newcommand{\sAppendix}[1]{%
     {\noindent\large\bfseries\appendixname\:: #1\par}%
     \sectionmark{#1}\vspace{\baselineskip}}

\def\CapitalUv{{\bf U}}
\def\CapitalVv{{\bf V}}

\allowdisplaybreaks

\begin{document}

\thispagestyle{empty}

\begin{center}

{\Large \bf
Thermal form-factor approach to dynamical correlation
functions of integrable lattice models}

\vspace{10mm}

{\large
Frank G\"{o}hmann,$^\dagger$ Michael Karbach,$^\dagger$ Andreas Kl\"umper,$^\dagger$
Karol K. Kozlowski,$^\ast$ and Junji Suzuki$^\ddagger$}\\[3.5ex]
$^\dagger$Fakult\"at f\"ur Mathematik und Naturwissenschaften,\\
Bergische Universit\"at Wuppertal,
42097 Wuppertal, Germany\\[1.0ex]
$^\ast$Univ Lyon, ENS de Lyon, Univ Claude Bernard,\\ CNRS,
Laboratoire de Physique, F-69342 Lyon, France\\[1.0ex]
$^\ddagger$Department of Physics, Faculty of Science, Shizuoka University,\\
Ohya 836, Suruga, Shizuoka, Japan

\vspace{25mm}

{\large {\bf Abstract}}

\end{center}

\begin{list}{}{\addtolength{\rightmargin}{9mm}
               \addtolength{\topsep}{-5mm}}
\item
We propose a method for calculating dynamical correlation functions
at finite temperature in integrable lattice models of Yang-Baxter
type. The method is based on an expansion of the correlation functions
as a series over matrix elements of a time-dependent quantum transfer
matrix rather than the Hamiltonian. In the infinite Trotter-number
limit the matrix elements become time independent and turn into the
thermal form factors studied previously in the context of static
correlation functions. We make this explicit with the example
of the XXZ model. We show how the form factors can be summed
utilizing certain auxiliary functions solving finite sets of
nonlinear integral equations. The case of the XX model is worked
out in more detail leading to a novel form-factor series
representation of the dynamical transverse two-point function.
\end{list}

\clearpage

\section{Introduction}
The goal of this work is to design a method for the calculation
of dynamical correlation functions at finite temperature in
integrable lattice models of Yang-Baxter type. This was attempted
before by K. Sakai \cite{Sakai07}, but the multiple-integral
formula he obtained turned out to be computationally inefficient.
Still, the basic idea in Sakai's work, which was to use the
`solution of the quantum inverse problem' \cite{KMT99a} twice,
for the usual transfer matrix and for the quantum transfer matrix,
seems very natural and is awaiting to be used in a more efficient
way. Here we combine this idea with the thermal form-factor expansion
introduced by two of the authors in \cite{DGK13a}. This leads
to a form-factor series of the same degree of complexity as in
the static case. In particular, only a single (rather than a
double) sum over excited states is involved.

We shall put some emphasis on the general formalism which applies
to a large class of integrable lattice models, namely to those
with an $R$-matrix which turns into the transposition matrix
for certain values of the spectral parameters. For all integrable
lattice models in this class we derive a form factor series for
their dynamical correlation functions at finite temperature in
the first part of this work. In order to evaluate the form-factor
series, which is the subject of the second part of this work,
the form factors should be known in a form which admits
to take the Trotter limit. Until recently this was only the case
for the form factors of the XXZ chain and of models directly related
to it as limiting cases. For the XXZ chain useful determinant
representations of the form factors \cite{IKMT99, KKMST09b,DGK13a}
were derived on the basis of Slavnov's scalar product formula
\cite{Slavnov89}. The representations obtained in \cite{DGK13a}
apply to the quantum transfer matrix and can be used to
take the Trotter limit. Rather recently determinant formulae
for form factors of local operators for the $gl(3)$ and
$gl(2|1)$ models were derived within an algebraic Bethe Ansatz
approach \cite{PRS15a,PRS15b, HLPRS16}, and important progress
toward a generalization to $gl(m|n)$ for generic values of $m$
and $n$ was made in \cite{HLPRS17pp}. These determinant formulae
may become the starting point for taking the Trotter limit
and for setting at work our novel form factor series in more
general higher-rank cases.

In order to make the form factor series efficient a partial
summation over classes of excitations seems necessary. Such
partial summation may be achieved by means of (multiple-)
integration over `auxiliary functions' for `higher-level Bethe
equations' which turns sums into integrals. This idea was developed
in \cite{DGKS15a} in the context of the usual transfer matrix
for ground state correlation functions of the XXZ model in the
antiferromagnetic massive regime and was first applied to thermal
form-factor series of the same model in \cite{DGKS15b,DGKS16b}.
In this work we suggest how to perform a similar partial
summation of the thermal form-factor series for the XXZ chain
in the dynamical case. This partial summation is quite different
from the partial summation employed in the analysis of the
long-time large-distance asymptotics of the form factor series
of the ground state correlation functions of the XXZ chain
in the critical regime \cite{KKMST11b,KKMST12}, which relied
on `restricted sums' rather than on contour integration. By way
of contrast the auxiliary function techniques developed in
\cite{DGKS15a,DGKS15b,DGKS16b} are closer in spirit to those of
\cite{GKS04a,KMST05a}.

Explicit limiting cases of our form factor series, in particular
the high- and low-temperature asymptotics, will be worked
out in future publications. In order to demonstrate that our
form-factor series can be efficient we focus here on the
special case of the XX model \cite{LSM61}. For the longitudinal
two-point functions our form factor series reduces to a
simple explicit formula equivalent to the classical result
of Niemeijer \cite{Niemeijer67,KHS70}. In the transverse
case we obtain a novel form factor series which seems to 
be rather promising for addressing some open questions concerning
the long-time and large distance asymptotics of the dynamical
two-point functions at finite temperature~\cite{IIKS93b,IIKS95,Jie98}.

\section{Foundations}
As far as the general formalism is concerned we shall work within
the setting of fundamental integrable models with unitary
$R$-matrix. In particular, no crossing symmetry will be required.
\subsection{Fundamental integrable lattice models}
Such models are defined in terms of their $R$-matrices
$R: {\mathbb C}^2 \mapsto \End \bigl( {\mathbb C}^d \otimes
{\mathbb C}^d \bigr)$ which are solutions of the Yang-Baxter
equation
\begin{equation} \label{ybe}
     R_{12} (\la,\m) R_{13} (\la,\n) R_{23} (\m,\n) =
        R_{23} (\m,\n) R_{13} (\la,\n) R_{12} (\la,\m) \epp
\end{equation}
The subscripts refer to the pairs of spaces in the triple
tensor product $\bigl({\mathbb C}^d\bigr)^{\otimes 3}$ on
which the corresponding $R$-matrix is acting nontrivially.
We shall assume that the $R$-matrix has the following
additional properties:
\begin{subequations}
\label{rprops}
\begin{align}
     & \text{regularity} & & R (\la, \la) = P \epc \label{reg} \\[1ex]
     & \text{unitarity} & & R_{12} (\la,\m) R_{21} (\m,\la) = \id \epc \label{uni} \\[1ex]
     & \text{symmetry} & & R^t (\la,\m) = R (\la,\m) \label{sym} \epp
\end{align}
\end{subequations}
Here the superscript $t$ denotes matrix transposition and $P$ is
the permutation matrix defined by $P (x \otimes y) = (y \otimes x)$
for all $x, y \in {\mathbb C}^d$.

With a given $R$-matrix, which has the above properties, we
associate an integrable lattice model in the standard way.
We define a (`row-to-row') monodromy matrix
\begin{equation}
     T_{\perp, a} (\la) = R_{aL} (\la,0) \dots R_{a1} (\la,0)
\end{equation}
on $L$ lattice sites and the corresponding transfer matrix
\begin{equation}
     t_\perp (\la) = \tr_a \{T_{\perp, a} (\la)\} \epp
\end{equation}
Then typically a constant $\ks \in {\mathbb C}$ exists such that,
for an appropriate choice of the parameters in the $R$-matrix,
\begin{equation} \label{hzerotrans}
     H_0 = \ks\, t_\perp' (0) t_\perp^{-1} (0)
\end{equation}
can be interpreted as a local lattice Hamiltonian, which is
Hermitian on $\End ({\mathbb C}^d)^{\otimes L}$. The locality is
obvious from the explicit expression
\begin{equation} \label{ham}
     H_0 = \ks \sum_{j=1}^L \6_\la (P R)_{j-1, j} (\la,0)\big|_{\la = 0} \epc
\end{equation}
where periodic boundary conditions, $(PR)_{0, 1} = (PR)_{L, 1}$,
are understood. Note that the constant $\ks$ depends on the
respective $R$-matrix.
\subsection{Transfer matrix realization of statistical operator
and time evolution operator}
In order to be able to calculate correlation functions we need
to realize the exponential of $H_0$ in terms of transfer matrices.
For this purpose we introduce
\begin{equation}
     \overline{t}_\perp (\la) = \tr_a \{ T^{-1}_{\perp, a} (\la) \}
        = \tr_a \{R_{1a} (0,\la) \dots R_{La} (0,\la) \} \epc
\end{equation}
where the second equation holds due to the unitarity condition
(\ref{uni}). Then it follows that
\begin{equation} \label{hzerotransbar}
     \overline{t}_\perp (0) = t_\perp^{-1} (0) \epc \qd
     H_0 = - \ks\, t_\perp (0) \overline{t}_\perp' (0) \epp
\end{equation}
Combining (\ref{hzerotrans}) and (\ref{hzerotransbar}) we obtain
\begin{equation} \label{exphinf}
     t_\perp \Bigl(- \frac{\ks}{NT}\Bigr)
        \overline{t}_\perp \Bigl(\frac{\ks}{NT}\Bigr) = 
	\id - \frac{2 H_0}{NT} + \CO \bigl(N^{-2}\bigr) \epp
\end{equation}
For every even $N$ let
\begin{equation} \label{defrho}
     \r_{N, L} (1/T) = \biggl( t_\perp \Bigl(- \frac{\ks}{NT}\Bigr)
        \overline{t}_\perp \Bigl(\frac{\ks}{NT}\Bigr) \biggr)^\frac{N}{2} \epp
\end{equation}
Then (\ref{exphinf}) implies that
\begin{equation} \label{euler}
     \lim_{N \rightarrow \infty} \r_{N, L} (1/T) = \re^{- H_0/T} \epp
\end{equation}
For finite $N$ the product of transfer matrices $\r_{N, L} (1/T)$
is an approximation to the statistical operator $\re^{- H_0/T}$,
where $T$ is the temperature. We shall call $N$ the Trotter number
and the limit $N \rightarrow \infty$ the Trotter limit.
\begin{remark}
In previous work we used to define $\r_{N, L} (1/T)$ with the opposite
order of factors. The order is irrelevant in (\ref{euler}), but
the present order turns out to be slightly more convenient, when
we consider dynamical correlation functions below.
\end{remark}
For the evaluation of correlation functions we will also have
to express the action of a local operator $x \in \End
({\mathbb C}^d)$ on the first site of our quantum chain in
terms of monodromy and transfer matrix. For this purpose we shall
employ the `solution of the quantum inverse problem' formula,
whose significance was first understood in \cite{KMT99a},
\begin{equation} \label{inversertr}
     x_1 = t_\perp (0) \tr_a \{ x_a T_{\perp, a}^{-1} (0) \}
         = \lim_{\e \rightarrow 0} t_\perp (- \e)
           \tr_a \{ x_a T_{\perp, a}^{-1} (\e) \} \epp
\end{equation}
Following \cite{Sakai07} we have introduced a regularization
parameter $\e$. The regularization is trivial for the row-to-row
transfer matrix. It becomes important only later when we apply
a variant of the above formula to the quantum transfer matrix
introduced in the next section.
\subsection{Quantum transfer matrix}
Now we introduce the central notion of our formalism, the
quantum transfer matrix \cite{Suzuki85}, which was previously
used in order to obtain efficient formulae for the free energy
per lattice site \cite{Kluemper93} and for static correlation
functions of integrable lattice models \cite{GKS04a}.
We shall see that, in a slightly generalized form, it
can be also used to study dynamical correlation functions
at finite temperature. This was already recognized in
\cite{Sakai07}, but the formula obtained by that time do
not seem to be efficient for a numerical or asymptotic
analysis.

We would like to include a simple class of external fields
which do not break the integrability of the model. They enter
the formalism in the following way. Let
$\hat \ph \in \End ({\mathbb C}^d)$ be a local operator, let
$\Th (\a) = \re^{\a \hat \ph}$, and assume that
\begin{equation} \label{groupinvariancer}
     [R_{12} (\la,\m), \Th_1 (\a) \Th_2 (\a)] = 0 \epp
\end{equation}
Then
\begin{equation} \label{grouplikeconserved}
     [t_\perp (\la), \Th_1 (\a) \dots \Th_L (\a)] =
     [\,\overline{t}_\perp (\la), \Th_1 (\a) \dots \Th_L (\a)] = 0 \epp
\end{equation}
Define
\begin{equation} \label{defph}
     \hat \PH = \sum_{j=1}^L \hat \ph_j \epp
\end{equation}
Then (\ref{grouplikeconserved}) implies that
\begin{equation}
     [H_0, \hat \PH] = 0 \epp
\end{equation}

Using $\Th(\a)$ we define the staggered, twisted and inhomogeneous
monodromy matrix acting on `vertical spaces' with site indices
$\overline{1}, \dots, \overline{2N + 2}$,
\begin{multline} \label{stagmon}
     T_a (\la|\a) \\ = \Th_a (\a) R_{\overline{2N+2},a}^{t_1} (\n_{2N+2},\la)
                    R_{a, \overline{2N+1}} (\la, \n_{2N+1}) \dots
                    R_{\overline{2},a}^{t_1} (\n_2,\la) R_{a,\overline{1}} (\la, \n_1)
		    \epp
\end{multline}
Here the superscript $t_1$ denotes transposition with respect
to the first space $R$ is acting on, and $\n_1, \dots, \n_{2N+2}$
are $2N+2$ arbitrary complex `inhomogeneity parameters'. The
corresponding transfer matrix
\begin{equation} \label{defqtm}
     t (\la|\a) = \tr_a \{T_a (\la|\a)\}
\end{equation}
is called the inhomogeneous quantum transfer matrix of the model.

The most important property of the staggered monodromy matrix
(\ref{stagmon}) is that it provides a representation of
the Yang-Baxter algebra,
\begin{equation} \label{yba}
     R_{ab} (\la,\m) T_a (\la|\a) T_b (\m|\a)
        = T_b (\m|\a) T_a (\la|\a) R_{ab} (\la,\m) \epp
\end{equation}
This follows from (\ref{ybe}), (\ref{groupinvariancer}) and
from the equation
\begin{equation} \label{ybet1}
     R_{ab} (\la,\m) R_{ja}^{t_1} (\n,\la) R_{jb}^{t_1} (\n,\m) =
        R_{jb}^{t_1} (\n,\m) R_{ja}^{t_1} (\n,\la) R_{ab} (\la,\m)
\end{equation}
which is obtained from the Yang-Baxter equation (\ref{ybe}) by
taking the transpose with respect to the first space, permuting
the indices and redefining the spectral parameters. In many cases
the Yang-Baxter algebra can be used in order to diagonalize the
quantum transfer matrix.

A crucial formula for our derivation of a form-factor series
for dynamical correlation functions at finite temperature is
a generalized form of the inversion formula \cite{KMT99a,GoKo00}.
\begin{lemma*}
Solution of the quantum inverse problem for the inhomogeneous
quantum transfer matrix. Let $j \in \{1, \dots, 2N+2\}$ be odd.
Then, for any $x \in \End \bigl({\mathbb C}^d\bigr)$,
\begin{multline} \label{soqip}
     x_j = t(\n_1|\a) t^{-1} (\n_2|\a) \dots t^{-1} (\n_{j-1}|\a) \\ \times
           \tr \{ x T(\n_j|\a) \} t^{-1} (\n_j|\a) t(\n_{j-1}|\a)
	   \dots t^{-1} (\n_1|\a) \epc
\end{multline}
provided that $t (\n_k|\a)$ is invertible for $k = 1, \dots, j$.
\end{lemma*}
A proof of this formula is given in Appendix~\ref{app:invqtm}.
In general invertibility will require that the inhomogeneity
parameters on odd and even lattice site are mutually distinct. Otherwise
for some $k$ and $\ell$ the determinant of $t (\n_k|\a)$ will
contain a factor $\det (P_{k, \ell}^{t_1}) = (\det (P^{t_1}))^{d^{2N}}$,
where the determinant on the left hand side is evaluated in
$({\mathbb C}^d)^{\otimes 2N + 2}$ and the
determinant on the right hand side is evaluated in ${\mathbb C}^d
\otimes {\mathbb C}^d$. But
\begin{equation}
     P^{t_1} (x \otimes y) = P^{t_1} (y \otimes x) \epc
\end{equation}
implying that $P^{t_1}$ has a nontrivial kernel and that
$\det (t (\n_k|\a)) = 0$. For our main example, the XXZ chain,
we shall provide sufficient conditions for $t (\n_k|\a)$ to
be invertible in Appendix~\ref{app:invshift}. Note that the
$\e$-regularization in (\ref{inversertr}) was introduced to
avoid invertibility problems in the intermediate calculations.
\subsection{Correlation functions}
We would like to calculate correlation functions of integrable
lattice models with Hamiltonian
\begin{equation} \label{fullham}
     H = H_0 - \a \hat \PH \epc
\end{equation}
where $H_0$ is defined in (\ref{ham}) and $\hat \PH$ in
(\ref{defph}). We restrict ourselves to the dynamical
two-point functions of two operators $x, y \in
\End \bigl({\mathbb C}^d\bigr)$ defined by
\begin{multline} \label{defcorrel}
     \<x_1 y_{m+1} (t)\>_T =
        \lim_{L \rightarrow \infty}
        \frac{\tr_{1, \dots, L}
	      \{ \re^{- H/T} x_1 \re^{\i t H} y_{m+1} \re^{-\i t H} \}}
             {\tr_{1, \dots, L} \{\re^{- H/T}\}} \\ =
        \lim_{L \rightarrow \infty}
        \frac{\tr_{1, \dots, L}
	      \{ \re^{- (1/T + \i t) H} x_1 \re^{\i t H} y_{m+1} \}}
             {\tr_{1, \dots, L} \{\re^{- H/T}\}} \epp
\end{multline}
Here we have denoted the spatial distance on the lattice by $m$
and the time by $t$. The indices $1, \dots, L$ in (\ref{defcorrel})
indicate that the traces are computed in $({\mathbb C}^d)^{\otimes L}$
which is the space of states of the Hamiltonian (\ref{fullham}).

\begin{figure}
\begin{center}
\includegraphics[width=.95\textwidth]{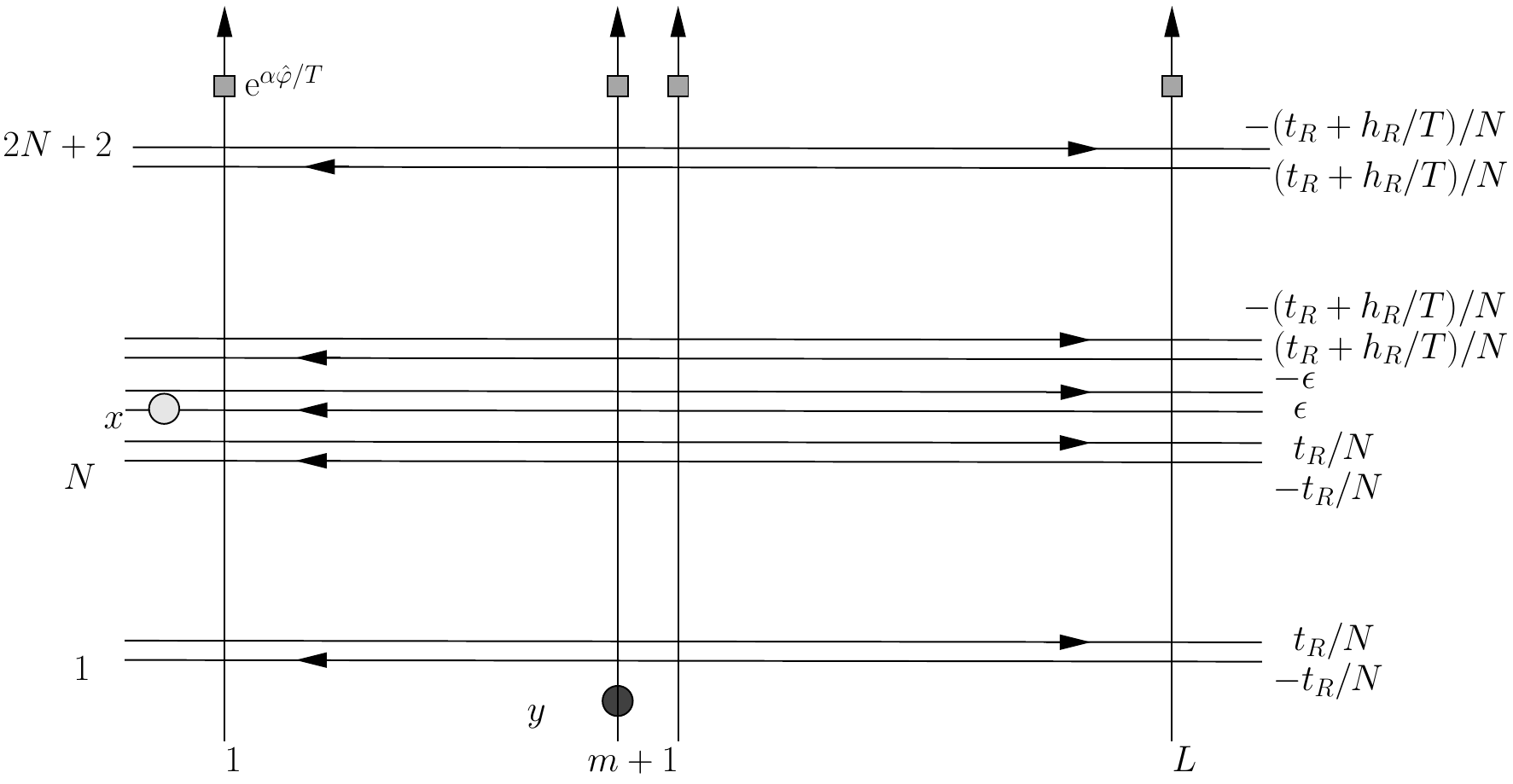}
\caption{\label{fig:dyn_correl_qft} Graphical representation
of (\ref{numcorrel}), the unnormalized finite Trotter number
approximant to the dynamical two-point function.}
\end{center}
\end{figure}
Our goal is to express the right hand side of (\ref{defcorrel})
in terms of the quantum transfer matrix (\ref{defqtm}) and
the entries of the corresponding monodromy matrix (\ref{stagmon}),
then to simplify the resulting expression using the spectral
decomposition of the quantum transfer matrix. Inserting
(\ref{defrho}) and (\ref{euler}) into the right hand side of
(\ref{defcorrel}) we obtain
\begin{multline} \label{correlrho}
     \<x_1 y_{m+1} (t)\>_T \\ =
        \lim_{L \rightarrow \infty}
        \lim_{N \rightarrow \infty}
        \frac{\tr_{1, \dots, L}
	      \{ \r_{N, L} (1/T + \i t) \re^{(1/T + \i t) \a \hat \PH} x_1
	      \re^{- \i t \a \hat \PH} \r_{N, L} (- \i t) y_{m+1} \}}
             {\tr_{1, \dots, L}
	     \{\re^{\a \hat \PH/T} \r_{N, L} (1/T + \i t) \r_{N, L} (- \i t) \}} \\ =
        \lim_{L \rightarrow \infty}
        \lim_{N \rightarrow \infty}
        \frac{\tr_{1, \dots, L}
	      \{ \re^{\a \hat \PH/T} \r_{N, L} (1/T + \i t)
	      \re^{\i t \a \hat \ph_1} x_1
	      \re^{- \i t \a \hat \ph_1} \r_{N, L} (- \i t) y_{m+1} \}}
             {\tr_{1, \dots, L}
	     \{\re^{\a \hat \PH/T} \r_{N, L} (1/T + \i t) \r_{N, L} (- \i t) \}} \epp
\end{multline}
We assume that the adjoint action of $\re^{\a \hat \ph}$
can be diagonalized. This is rather natural, since the operator
$\hat \ph$ is typically a Cartan element of a Lie algebra
acting as a local symmetry. Then, without loss of generality,
one may assume that $x$ is an eigenvector under the adjoint
action of $\hat \ph$, implying that
\begin{equation} \label{adjointev}
     \re^{\i t \a \hat \ph_1} x_1 \re^{- \i t \a \hat \ph_1}
        = \re^{\i t \a s(x)} x_1 \epc
\end{equation}
where $s(x)$ is the eigenvalue corresponding to $x$. We first of all
concentrate on the numerator on the right hand side of (\ref{correlrho}).
Inserting (\ref{adjointev}) and using (\ref{inversertr}) we find
\begin{multline} \label{numcorrel}
        \tr_{1, \dots, L}
	   \bigl\{ \re^{\a \hat \PH/T} \r_{N, L} (1/T + \i t)
	   \re^{\i t \a \hat \ph_1} x_1
	   \re^{- \i t \a \hat \ph_1} \r_{N, L} (- \i t) y_{m+1} \bigr\} =
	   \re^{\i t \a s(x)} \\ \times
	   \lim_{\e \rightarrow 0}
           \tr_{1, \dots, L}
	   \bigl\{ \re^{\a \hat \PH/T} \r_{N, L} (1/T + \i t)
	   t_\perp (- \e) \tr_a \{ x_a T_{\perp, a}^{-1} (\e) \}
	   \r_{N, L} (- \i t) y_{m+1} \bigr\} \epp
\end{multline}

This is now of a form that allows us to write it in terms of the
quantum transfer matrix and its monodromy matrix. The easiest
way to proceed is to represent the right hand side graphically
(see Fig.~\ref{fig:dyn_correl_qft}) and re-express it in terms of
column-to-column rather than row-to-row monodromy matrices. The
column-to-column monodromy matrix at hand is a special case of
the staggered monodromy matrix defined in (\ref{stagmon}). The
inhomogeneities can, for instance, be fixed to
\begin{equation} \label{trotterdecomp}
    \n_{2k-1} = - \n_{2k} =
                  \begin{cases}
		     - \frac{\tsuz}{N} & k = 1, \dots, \frac{N}{2} \\
		     \e & k = \frac{N}{2} + 1 \\
		     \frac{\tsuz + \ks /T}{N} & k = \frac{N}{2} + 2, \dots, N+1 \epc
                  \end{cases}
\end{equation}
where
\begin{equation} \label{tsuz}
    \tsuz = \i \ks t \epp
\end{equation}
Then
\begin{align} \label{numcorrelqtm}
     \tr_{1, \dots, L} &
        \bigl\{ \re^{\a \hat \PH/T} \r_{N, L} (1/T + \i t)
	t_\perp (- \e) \tr_a \{ x_a T_{\perp, a}^{-1} (\e) \}
	\r_{N, L} (- \i t) y_{m+1} \bigr\} \notag \\[1ex] =
     & \tr_{\overline{1}, \dots, \overline{2N+2}}
        \bigl\{ x_{\overline{N+1}} t^m (0|\k) \tr \{ y T(0|\k) \}
	        t^{L-m-1} (0|\k) \bigr\} \notag \\[1ex] =
     & \tr_{\overline{1}, \dots, \overline{2N+2}}
        \biggl\{ \biggl[ t \Bigl(- \frac{\tsuz}{N} \Big|\k \Bigr)
	        t^{-1} \Bigl(\frac{\tsuz}{N} \Big|\k \Bigr) \biggr]^\frac{N}{2}
		\tr \{x T(\e|\k)\} t^{-1} (\e|\k) \notag \\ & \qd \times
                \biggl[ t^{-1} \Bigl(- \frac{\tsuz}{N} \Big|\k \Bigr)
	        t \Bigl(\frac{\tsuz}{N} \Big|\k \Bigr) \biggr]^\frac{N}{2}
		t^m (0|\k) \tr \{y T(0|\k)\} t^{L-m-1} (0|\k) \biggr\} \epc
\end{align}
where we wrote $\k = \a/T$ for short and used (\ref{soqip}) in
the second equation. We reinsert (\ref{numcorrelqtm}) back
into (\ref{numcorrel}), (\ref{correlrho}) and take the limit
$L \rightarrow \infty$ first, exploiting the fact that there is
a single dominant eigenvalue $\La_0$ with eigenvector $|\Ps_0\>$
of the quantum transfer matrix. The interchangeability of the
limits was discussed in \cite{Suzuki85}, albeit for a slightly
differently defined quantum transfer matrix. The existence of a
dominant eigenvalue at finite Trotter number is at least clear
at high enough temperature. We obtain the following representation
for the two-point functions in the thermodynamic limit,
\begin{multline} \label{correlpsipsi}
     \<x_1 y_{m+1} (t)\>_T =
        \lim_{N \rightarrow \infty}
        \lim_{\e \rightarrow 0} \re^{\i t \a s(x)}
        \biggl( \frac{\La_0 \bigl(- \frac{\tsuz}{N}\big|\k \bigr)}
                     {\La_0 \bigl(\frac{\tsuz}{N}\big|\k \bigr)} \biggr)^\frac{N}{2}
		     \\ \times
        \frac{\<\Ps_0| X(\e|\k) t^{-1} (\e|\k)
                \Bigl[ t^{-1} \bigl(- \frac{\tsuz}{N} \big|\k \bigr)
	        t \bigl(\frac{\tsuz}{N} \big|\k \bigr) \Bigr]^\frac{N}{2}
		t^m (0|\k) Y(\e|\k)|\Ps_0\>}
             {\La_0 (\e|\k) \La_0^m (0|\k) \<\Ps_0|\Ps_0\>} \epp
\end{multline}
Here we have introduced the notation
\begin{equation}
	X(\la|\k) = \tr \{x T(\la|\k)\} \epc \qd
	Y(\la|\k) = \tr \{y T(\la|\k)\} \epp
\end{equation}
We also took the liberty to change the spectral parameter
from zero to $\e$ on one of the vertical lines by replacing
$Y(0|\k)$ by $Y(\e|\k)$ and by performing a similar replacement
in the denominator. This will produce more symmetric and slightly
more convenient expressions later on.

In order to be able to deal with the Trotter limit $N \rightarrow \infty$
we insert a complete set of eigenstates $|\Ps_n\>$ of $t(\la|\k)$ with
corresponding eigenvalues $\La_n (\la|\k)$. This brings us to our main
result.
\begin{theorem*}
With the definitions above the dynamical two-point functions
of two local operators $x$ and $y$ have the form-factor series
expansion
\begin{multline} \label{ffseriesnfinite}
     \<x_1 y_{m+1} (t)\>_T =
        \lim_{N \rightarrow \infty}
        \lim_{\e \rightarrow 0}
	\re^{\i t \a s(x)} \sum_n
        \frac{\<\Ps_0|X(\e|\k)|\Ps_n\>\<\Ps_n|Y(\e|\k)|\Ps_0\>}
             {\La_n (\e|\k) \<\Ps_0|\Ps_0\> \La_0 (\e|\k) \<\Ps_n|\Ps_n\>} \\ \times
        \biggl( \frac{\La_n (0|\k)}{\La_0 (0|\k)} \biggr)^m
        \biggl( \frac{\La_n \bigl(\frac{\tsuz}{N}\big|\k \bigr)
	              \La_0 \bigl(- \frac{\tsuz}{N}\big|\k \bigr)}
                     {\La_0 \bigl(\frac{\tsuz}{N}\big|\k \bigr)
                      \La_n \bigl(- \frac{\tsuz}{N}\big|\k \bigr)} \biggr)^\frac{N}{2}
		      \epp
\end{multline}
\end{theorem*}
\begin{Remark}
Setting $t = 0$ we formally recover the known form-factor series expansion
of the static correlation functions of integrable lattice models
\cite{DGK13a}.
\end{Remark}
\begin{Remark}
The Trotter limit can be performed for the individual terms occurring
under the sum,
\begin{align} \label{defamps}
     & A_n = \lim_{N \rightarrow \infty}
           \lim_{\e \rightarrow 0}
           \frac{\<\Ps_0|X(\e|\k)|\Ps_n\>\<\Ps_n|Y(\e|\k)|\Ps_0\>}
	        {\La_n (\e|\k) \<\Ps_0|\Ps_0\> \La_0 (\e|\k) \<\Ps_n|\Ps_n\>}
		\epc \\[1ex]
     & \r_n (\la) =
           \lim_{N \rightarrow \infty}
           \lim_{\e \rightarrow 0}
           \frac{\La_n (\la |\k)}{\La_0 (\la |\k)} \epp
\end{align}
With this we obtain the formal series expansion
\begin{equation} \label{ffseries}
     \<x_1 y_{m+1} (t)\>_T = \sum_n A_n \r_n^m (0)
                             \re^{\i t \{ \a s(x) + \ks \r_n' (0)/\r_n (0)\}}
\end{equation}
from (\ref{ffseriesnfinite}). We would like to emphasize, however,
that the Trotter limit in (\ref{ffseriesnfinite}) has to be dealt
with with care, taking into account the peculiarities of the given
model under consideration. The specific example of the XXZ chain
will be considered in the following section.
\end{Remark}
\begin{Remark}
Another way of looking at (\ref{ffseries}) is by introducing
the correlation lengths $\x_n$ and the phase velocity $v_n$,
setting
\begin{equation}
     \r_n (\la) = \re^{- \frac{1}{\x_n (\la)}} \epc \qd
     v_n = \frac{\x_n'(0)}{\x_n (0)} \epp
\end{equation}
Then
\begin{equation}
     \<x_1 y_{m+1} (t)\>_T =
        \sum_n A_n
	\exp \biggl\{- \frac{m - v_n \tsuz}{\x_n (0)}
	             + \i t \a s(x) \biggr\} \epp
\end{equation}
\end{Remark}
\begin{Remark}
In our derivation we can easily change the Trotter
decomposition (\ref{trotterdecomp}). This allows us to
modify the Hamiltonian and to replace it by any linear
combination of local conserved quantities generated
by the row-to-row transfer matrix of the model \cite{KlSa02}.
\end{Remark}
\section{The XXZ chain as an example}
\subsection{Hamiltonian and R-matrix}
In this section we shall explore some of the features of our
approach using the example of the XXZ chain. Only for this
most elementary model the amplitudes occurring in the
form-factor expansion of the two-point functions have been
worked out in sufficient detail.

The Hamiltonian of the spin-$\2$ XXZ chain in a magnetic field
of strength $h$ is defined by the local action of Pauli matrices
$\s^\a$, $\a = x, y, z$, on a chain of spins on $L$ lattice sites,
\begin{equation} \label{hxxz}
     H_{XXZ} = J \sum_{j = 1}^L \Bigl( \s_{j-1}^x \s_j^x + \s_{j-1}^y \s_j^y
                 + \D \bigl( \s_{j-1}^z \s_j^z - 1 \bigr) \Bigr)
		 - \frac{h}{2} \sum_{j=1}^L \s_j^z \epp
\end{equation}
Here $\D = (q + q^{-1})/2$ is the anisotropy parameter and $J > 0$ is
the strength of the exchange interaction. In the following
we shall restrict ourselves to $q = \re^{- \i \g}$, $\g \in (0,\p/2]$
for simplicity, implying that $0 \le \D < 1$.

The Hamiltonian $H_{XXZ}$ can be obtained from the $R$-matrix
\begin{subequations}
\label{rxxz}
\begin{align}
     R(\la,\m) & = \begin{pmatrix}
                    1 & 0 & 0 & 0 \\
		    0 & b(\la, \m) & c(\la, \m) & 0 \\
		    0 & c(\la, \m) & b(\la, \m) & 0 \\
		    0 & 0 & 0 & 1
		   \end{pmatrix} \epc
		   \displaybreak[0] \\[2ex] \label{defbc}
     b(\la, \m) & = \frac{\sh(\la - \m)}{\sh(\la - \m - \i \g)} \epc \qd
     c(\la, \m) = \frac{\sh(- \i \g)}{\sh(\la - \m - \i \g)}
\end{align}
\end{subequations}
which is regular, unitary and symmetric (see (\ref{rprops})).
It has a $U(1)$ symmetry of the form (\ref{groupinvariancer})
with $\Th(\a) = q^{\a \s^z}$ or $\hat \ph = - \i \g \s^z$.%
\footnote{Note that $\s^z$ and $\s^\pm$ are eigenvectors under
the adjoint action of $\hat \ph$ with eigenvalues $s(\s^z) = 0$
and $s(\s^\pm) = \mp 2 \i \g$.} Thus, we may construct $H$ as
in (\ref{fullham}). This Hamiltonian turns into $H_{XXZ}$ if
we choose
\begin{equation}
     \ks = - 2 \i J \sin (\g) \epc \qd \a = \frac{\i h}{2 \g} \epp
\end{equation}
Then $\k = \i h/2 \g T$ and $\tsuz = 2 J \sin(\g) t$.

For the XXZ chain we can also calculate the determinants of
the `inhomogeneous shift operators' $t(\n_j|\a)$ and find that
they vanish if $\n_{2j-1} = \n_{2k}$ for some $j, k \in
\{1, \dots, N+1\}$. This was the reason why we have introduced
the $\e$-regularization. A set of sufficient conditions for
the invertibility of the $t(\n_j|\a)$ is that the inhomogeneity
parameters on odd and even sites be mutually distinct and
that $|\n_j| < \g/2$ for $j = 1, \dots, 2N + 2$ (see
Appendix~\ref{app:invshift}).
\subsection{Algebraic Bethe Ansatz}
In order to set our form-factor series (\ref{ffseriesnfinite})
at work we first of all need to know the eigenvalues and eigenvectors
of the quantum transfer matrix. For the XXZ chain the eigenvectors
can be constructed by means of the algebraic Bethe Ansatz.
This has been described at many instances\footnote{The completeness
of the system of eigenvectors obtained through the Bethe Ansatz
is a separate issue. It was shown in \cite{TaVa95} that all
solutions to the Bethe equations with pairwise distinct Bethe roots
provide a complete set of eigenvectors of the quantum transfer
matrix if the magnetic field and the inhomogeneities are generic.
In such a situation the Bethe vectors $|\Psi_n\>$ are all
well-defined and, in particular, have non-vanishing `norm'
$\<\Psi_n|\Psi_n\>$.  For simplicity we shall work with the
specific choice of the inhomogeneities $\nu_k$ as given in
\eqref{trotterdecomp}. If any problem related to completeness or
vanishing of the norm  $\<\Psi_n|\Psi_n\>$ should arise, it would
be enough to slightly perturb the $\nu_k$ in the intermediate
calculations and then send them to the values \eqref{trotterdecomp}
in the end, on the level of the final formula, where the limit
is already regular.} (see e.g.\ \cite{KBIBo}). Here we only have
to adapt the known result to our conventions. We consider the
inhomogeneous monodromy matrix (\ref{stagmon}) and write it as
a $2 \times 2$ matrix in `auxiliary space' $a$,
\begin{equation}
     T_a (\la|\k) = \begin{pmatrix}
                    A(\la) & B(\la) \\ C(\la) & D(\la)
                 \end{pmatrix}_a \epp
\end{equation}
This defines the operators $A(\la), \dots, D(\la)$. The
operators $B(\la)$ generate the eigenstates of the quantum
transfer matrix $t(\la|\k) = A(\la) + D(\la)$ by acting
on a pseudo vacuum $|0\>$ defined by $C(\la) |0\> = 0$.
In our case the pseudo vacuum is
\begin{equation}
     |0\> = \biggl( \binom{1}{0} \otimes \binom{0}{1} \biggr)^{\otimes (N+1)} \epp
\end{equation}
The pseudo vacuum is an eigenvector of the operators
$A(\la)$, with eigenvalue $a(\la)$, and $D(\la)$, with
eigenvalue $d(\la)$. The eigenvalues are readily calculated
using the explicit form of the $R$-matrix and its transposed
with respect to the first space,
\begin{equation}
     a(\la) = q^\k \prod_{k=1}^{N+1} b(\n_{2k},\la) \epc \qd
     d(\la) = q^{-\k} \prod_{k=1}^{N+1} b(\la,\n_{2k-1}) \epp
\end{equation}

In the context of the quantum transfer matrix formalism it
has turned out to be useful \cite{Kluemper93} to describe the
Bethe Ansatz solution in terms of certain auxiliary functions.
For $M = 0, \dots, 2N+2$ define a family of functions
\begin{equation} \label{defaux}
     \fa \bigl(\la|\{\la_k\}_{k=1}^M, \k \bigr) = \frac{d(\la)}{a(\la)}
                 \prod_{k=1}^M \frac{\sh(\la - \la_k - \i \g)}
                                    {\sh(\la - \la_k + \i \g)}
\end{equation}
depending meromorphically on $M$ complex parameters $\la_j$.
The equations
\begin{equation} \label{baes}
     \fa \bigl(\la_j|\{\la_k\}_{k=1}^M, \k \bigr) = - 1 \epc \qd j = 1, \dots, M
\end{equation}
are called the Bethe Ansatz equations. Their solutions
$\{\la_j^{(n)}\}_{j=1}^M$ are sets of `Bethe roots'. We have supplied
a superscript `$(n)$' to distinguish the different solutions at fixed
$M$. With every set of Bethe roots we associate its corresponding
auxiliary function
\begin{equation}
     \fa_n (\la|\k) = \fa \bigl(\la|\{\la_k^{(n)}\}_{k=1}^M, \k \bigr) \epp
\end{equation}

Sets of Bethe roots $\{\la_j^{(n)}\}_{j=1}^M$ parameterize the
solutions of the eigenvalue problem of the quantum transfer
matrix. The eigenvalues can be written as
\begin{equation} \label{evasxxz}
     \La_n (\la|\k) =
        a(\la) \prod_{j=1}^M
	   \frac{\sh(\la - \la_j^{(n)} + \i \g)}{\sh(\la - \la_j^{(n)})} +
        d(\la) \prod_{j=1}^M
	   \frac{\sh(\la - \la_j^{(n)} - \i \g)}{\sh(\la - \la_j^{(n)})} \epp
\end{equation}
The eigenvectors and their `duals' take the form
\begin{equation} \label{evs}
     |\Ps_n\> 
        = B(\la_1^{(n)}) \dots B(\la_M^{(n)}) |0\> \epc \qd
     \<\Ps_n| 
        = \bigl\<0\big| C(\la_1^{(n)}) \dots C(\la_M^{(n)}) \epp
\end{equation}
\subsection{Nonlinear integral equations for the auxiliary functions
and integral representations of the eigenvalue ratios}
Let us write the auxiliary function for the general inhomogeneous quantum
transfer matrix of the XXZ chain explicitly,
\begin{multline} \label{auxexp}
     \fa_n (\la|\k) = \\ (-1)^s q^{- 2\k} \biggl[
                 \prod_{k=1}^{N+1} \frac{\sh(\la - \n_{2k-1})}{\sh(\la - \n_{2k})}
                    \frac{\sh(\i \g + \la - \n_{2k})}{\sh(\i \g - \la + \n_{2k-1})} \biggr]
                 \prod_{k=1}^M \frac{\sh(\i \g - \la + \la_k^{(n)})}
                                    {\sh(\i \g + \la - \la_k^{(n)})} \epc
\end{multline}
where
\begin{equation} \label{pseudospin}
     s = N + 1 - M
\end{equation}
is the (pseudo) spin of the excitation.

The reasoning that brings us a rough understanding of the structure
of the solutions of the Bethe Ansatz equations (\ref{baes}) is the same
as in the staggered case. The Bethe roots $\la_j^{(n)}$ are located on
the level curve $|\fa_n (\la|\k)| = 1$. If the $\n_j$ are mutually distinct
and of order $1/N$, then $\fa_n(\la|\k)$ has $N+1$ simple poles at $\n_{2k}$
and $N+1$ simple zeros at $\n_{2k-1}$, both close to $\la = 0$. If we
group the poles and zeros in close-by pairs and join each pair by
a straight line, then the function $|\fa_n(\la|\k)|$ takes on the value
$1$ at some point on each of these lines. The contour $|\fa_n(\la|\k)| = 1$
is connecting these points. Because of the many poles and zeros
the phase of $\fa_n(\la|\k)$ strongly varies along this contour, meaning that
close to zero there must be many points on the contour, where $\fa_n(\la|\k)
= - 1$.

Let us first focus on the case $M = N + 1$. We shall look for a
special solution $\{\la_j^{(0)}\}_{j=1}^{N+1}$ to the Bethe equations
for which all $\la_j^{(0)}$, $j = 1, \dots, N + 1$, are of the form
$\la_j^{(0)}= {\cal O} \bigl(j/(TN)\bigr)$. For such a solution
\begin{equation}
     \prod_{k=1}^{N+1}
        \frac{\sh(\la - \n_{2k} + \i \g)}{\sh(\la - \n_{2k-1} - \i \g)}
	\frac{\sh(\la - \la_k^{(0)} - \i \g)}{\sh(\la - \la_k^{(0)} + \i \g)} =
	c(\la) = 1 + \re^{\CO (1/T)}
\end{equation}
as long as $\la$ stays away from the poles close to $\pm \i \g$.
Setting $z = \re^{2 \la}$ and $z_k = \re^{2 \n_k}$ we see that
the equation $\fa_0 (\la|\k) = - 1$ is equivalent to
\begin{equation}
     p(z) = c(\la(z)) \, \re^{- \frac{h + \ks}T} \prod_{k=1}^{N+1} (z - z_{2k-1}) +
            \prod_{k=1}^{N+1} (z - z_{2k}) = 0 \epc
\end{equation}
if
\begin{equation} \label{sumisbeta}
     \sum_{k=1}^{N+1} (\n_{2k-1} - \n_{2k}) = \frac{\ks}{T} \epp
\end{equation}
Note that our Trotter decomposition (\ref{trotterdecomp}) satisfies
the latter condition for $\e \rightarrow 0$. Sticking with this
example we see that for $|\ks/T|, |\tsuz| \rightarrow 0$ we have
$z_k \rightarrow 1$ which should define the high-temperature regime
also in the general inhomogeneous case, whence
\begin{equation}
     p(z) \rightarrow (q^{-2 \k} + 1)(z - 1)^{N+1}
\end{equation}
in the high-temperature limit. Thus, we have an $(N+1)$-fold zero
at $z = 1$, which, for small $|\ks/T|$, $|\tsuz|$, is split into $N+1$ zeros
close to $z = 1$, corresponding to $N+1$ Bethe roots $\la_j^{(0)}$ close to
zero. These form a self-consistent high-temperature solution
of the Bethe Ansatz equations. Inserting the solution back into
(\ref{auxexp}) we see that the other zeros of $1 + \fa_0$ must be
located close to the poles at $\pm \i \g$. More precisely, we expect
$N+1$ of them close to $\i \g$ and $N+1$ close to $- \i \g$.

Defining the canonical contour ${\cal C}_0$ in the usual way
as $(- \i \g/2 + \i \de - \infty, - \i \g/2 + \i \de + \infty) \cup
(\i \g/2 - \i \de + \infty, \i \g/2 - \i \de - \infty)$, where
$\de > 0$ is small, we observe that for $N$ large enough the
Bethe roots of the above solution are inside the contour,
while all other zeros of $1 + \fa_0$ remain outside. This is enough
information to derive a nonlinear integral equation for the
auxiliary function associated with this specific solution of
the Bethe Ansatz equations.

Note that for our choice of parameters the function
$\ln (\sh(\i \g - \la + \m)/\sh(\i \g + \la - \m))$, where `$\ln$'
denotes the principal branch of the logarithm, is analytic inside
${\cal C}_0$. We fix a point $x_0 \in {\cal C}_0$ and, for
every $\la \in {\cal C}_0$, define a contour ${\cal C}_{x_0}^\la$
starting at $x_0$ and running along ${\cal C}_0$ up to the
point $\la$. This enables us to define
\begin{equation} \label{deflogoneplusa}
     \ln (1 + \fa_0) (\la|\k)
        = \int_{{\cal C}_{x_0}^\la} \rd \m \:
	  \frac{\fa_0' (\m|\k)}{1 + \fa_0 (\m|\k)} \epp
\end{equation}
It follows that
\begin{multline} \label{derivenlie}
     \int_{{\cal C}_0} \frac{\rd \m}{2 \p \i}
        \ln \biggl( \frac{\sh(\i \g - \la + \m)}{\sh(\i \g + \la - \m)} \biggr)
	\frac{\fa_0' (\m|\k)}{1 + \fa_0 (\m|\k)} \\[1ex]
     = - \int_{{\cal C}_0} \frac{\rd \m}{2 \p \i} K(\la - \m)
         \ln (1 + \fa_0) (\m|\k) =
         \ln \bigl(\fa_0 (\la|\k)\bigr) - 2 \i \g \k + \ks \re_N (\la)/T \epc
\end{multline}
where
\begin{subequations}
\begin{align}
     & K(\la) = \cth(\la + \i \g) - \cth(\la - \i \g) \epc \\[1ex]
     & \re_N (\la) = \frac{T}{\ks} \sum_{k=1}^{N+1}
	\ln \biggl( \frac{\sh(\la - \n_{2k})}{\sh(\la - \n_{2k-1})}
	            \frac{\sh(\la - \n_{2k-1} - \i \g)}{\sh(\la - \n_{2k} - \i \g)}
		    \biggr) \epp \label{defen}
\end{align}
\end{subequations}
In (\ref{derivenlie}) we have used the fact that $1 + \fa_0$ has as many
poles as zeros inside ${\cal C}_0$ when we performed the partial
integration in the first equation. In the second equation we have
used our knowledge about the location of the poles and zeros of $1 + \fa_0$
and the explicit representation (\ref{auxexp}) of the auxiliary function.
Equation (\ref{derivenlie}) can be read as a nonlinear integral equation
for the auxiliary function $\fa_0$,
\begin{equation} \label{nliedomfinn}
     \ln \bigl(\fa_0 (\la|\k)\bigr) = 2 \i \g \k - \ks \re_N (\la)/T
        - \int_{{\cal C}_0} \frac{\rd \m}{2 \p \i} K(\la - \m)
	                    \ln (1 + \fa_0) (\m|\k) \epp
\end{equation}

This equation is an inhomogeneous version of the well-known 
\cite{Kluemper93,GKS04a} nonlinear integral equation for the auxiliary
function of the dominant eigenvalue of the quantum transfer matrix.%
\footnote{For more general inhomogeneous Trotter decompositions
see \cite{KlSa02}.} Our derivation shows that it holds for any
inhomogeneous Trotter decomposition if $|\n_j|$, $j = 1, \dots,
2N + 2$, is small enough uniformly in $j$. It holds in particular
for our Trotter decomposition (\ref{trotterdecomp}) for finite
but small enough $|\ks/T|$, $|\tsuz|$. Our experience with the
numerical solution of (\ref{nliedomfinn}) suggests that it
has a unique solution. In forthcoming work we shall show that
this is indeed the case if the $|\n_j|$ are small enough. This
implies, at least for $M = N + 1$, that all other solutions to
the Bethe equations, which correspond to `excited states' must
contain roots which are away from the origin if the $|\n_j|$ are
small.

Numerical studies also suggest that the contour ${\cal C}_0$ is
independent of the Trotter number implying that the Trotter
limit $N \rightarrow \infty$ only affects the term $\re_N (\la)$.
For our Trotter decomposition (\ref{trotterdecomp}) we also
have to send $\e$ to zero. Then
\begin{equation} \label{bareenergy}
     \lim_{\substack{N \rightarrow \infty\\ \e \rightarrow 0}} \re_N (\la) =
     \re (\la) = \cth(\la) - \cth(\la - \i \g)
\end{equation}
which is the bare energy function. With this the nonlinear integral
equation for the auxiliary function of the dominant state in
the Trotter limit $\fatl_0$ takes its familiar form \cite{Kluemper93,%
GKS04a}
\begin{equation} \label{nliedomi}
     \ln \bigl(\fatl_0 (\la|\k)\bigr) = 2 \i \g \k - \ks \re(\la)/T
        - \int_{{\cal C}_0} \frac{\rd \m}{2 \p \i} K(\la - \m)
	                    \ln (1 + \fatl_0) (\m|\k) \epp
\end{equation}
The function $\fatl_0$ determines the thermodynamic properties and
the static temperature dependent correlation functions of the XXZ
chain at all finite temperatures \cite{Kluemper93,GKS04a} as it
parameterizes the integral representations of the dominant eigenvalue
and of the reduced density matrix of the model. The amazing fact,
which we would like to emphasize and which was already observed
by Sakai \cite{Sakai07}, is that no dependence on time $t$ has remained
in the Trotter limit.

The thermal form factors and eigenvalue ratios in
(\ref{ffseriesnfinite}) are parameterized by the auxiliary functions
connected with excited states of the quantum transfer matrix. For
these auxiliary functions several alternative descriptions are available.
The formally simplest one uses equivalence classes of contours
${\cal C}_n$ in order to classify the excitations (see e.g.\
\cite{DGK14a}). For any given auxiliary function $\fa_n$ there exists
a contour ${\cal C}_n$ which encircles all the Bethe roots, but no other
zeros of $1 + \fa_n$ and no other poles of this function than those at
$\n_{2k}$, $k = 1, \dots, N + 1$. We shall assume that we can shape
the contour ${\cal C}_n$ in such a way that $\la - \m \pm \i \g$ remains
outside for $\la, \m \in {\cal C}_n$. For simplicity we also assume
that ${\cal C}_n$ contains all Bethe roots of the dominant state and
no additional pole or zero of $1 + \fa_0$ as compared to ${\cal C}_0$.

With these prerequisites the only difference in the derivation
of a nonlinear integral equation is that the partial integration
performed in (\ref{derivenlie}) produces additional boundary terms
due to the fact that for an excitation with $M$ Bethe roots
\begin{equation} \label{cnmonodromy}
     \int_{{\cal C}_n} \frac{\rd \m}{2 \p \i}
        \frac{\fa_n' (\m|\k)}{1 + \fa_n (\m|\k)} = M - N - 1 = - s \epp
\end{equation}
As in case of the dominant state we fix a point $x_n$ on ${\cal C}_n$
and define the sub-contour ${\cal C}_{x_n}^\la$ connecting $x_n$
and $\la \in {\cal C}_n$ in positive direction along ${\cal C}_n$.
Then $\ln (1 + \fa_n) (\la|\k)$ can be defined in analogy with
(\ref{deflogoneplusa}). Using this determination of the logarithm
partial integration gives us
\begin{multline} \label{derivenlieiexcite}
     \int_{{\cal C}_n} \frac{\rd \m}{2 \p \i}
        \ln \biggl( \frac{\sh(\i \g - \la + \m)}{\sh(\i \g + \la - \m)} \biggr)
	\frac{\fa_n' (\m|\k)}{1 + \fa_n (\m|\k)} = \\[1ex]
	- \int_{{\cal C}_n} \frac{\rd \m}{2 \p \i} K(\la - \m) \ln (1 + \fa_n) (\m|\k)
	- s \ln \biggl( \frac{\sh(\i \g - \la + x_n)}
	                     {\sh(\i \g + \la - x_n)} \biggr) \epp
\end{multline}

Assuming that the contour ${\cal C}_n$ is such that we can send
$\Re x_n \rightarrow - \infty$ we obtain the nonlinear integral
equation
\begin{equation} \label{nlieexc}
     \ln \bigl(\fa_n (\la|\k)\bigr) = \i \p s + 2 \i \g (\k + s) - \ks \re_N (\la)/T
        - \int_{{\cal C}_n} \frac{\rd \m}{2 \p \i} K(\la - \m)
	                    \ln (1 + \fa_n) (\m|\k)
\end{equation}
for the excited states of the quantum transfer matrix in a form
parameterized by the contour~${\cal C}_n$. Solutions of (\ref{nlieexc})
are classified by equivalence classes of contours ${\cal C}_n$,
two contours being equivalent if they admit the same solution
$\fa_n (\cdot|\k)$. Another form of equations is obtained if we
deform all contours for a given value of $s$ to some reference
contour ${\cal C}_{0, s}$. In the process of the deformation the
contour will cross branch points of $\ln (1 + \fa_n)$ which will
appear as `particle and hole parameters' in the additional driving
terms generated on the right hand side of (\ref{nlieexc}).
This possibility will be further elaborated in the next section,
where we discuss the summation of the form-factor series in the
general case.

About the Trotter limit of the functions $\fa_n$ the same can be
said as in case of the dominant state. Our experience tells
us that the contour ${\cal C}_n$ becomes independent of $N$ in
the Trotter limit. The only part which depends on $N$ and $\e$
is the function $\re_N$. In the limit $N \rightarrow \infty$,
$\e \rightarrow 0$ it turns into the bare energy (\ref{bareenergy}),
and no dependence on $t$ is remaining. Now all functions
appearing in the form-factor expansion are parameterized by the
auxiliary functions $\fa_0$ and $\fa_n$ and by the corresponding
contours. Their dependence on $\fa_0$ and $\fa_n$ is always the
same independent of the underlying Trotter decomposition. If we
choose our Trotter decomposition (\ref{trotterdecomp}) they will
all be time dependent, but the time dependence will vanish in the
Trotter limit. At this point we understand that we can fully resort
to the results of \cite{DGK13a} for the eigenvalue ratios and
the amplitudes $A_n$ in the form-factor series (\ref{ffseriesnfinite}).

Let us recall the expressions for the eigenvalue ratios. Following
\cite{BoGo09} we shall consider ratios of eigenvalues with different
values of the magnetic field. We will use the function
\begin{equation}
     z_n (\la|\k, \k') =
        \frac{\ln (1 + \fa_0) (\la|\k) - \ln (1 + \fa_n) (\la|\k')}{2 \pi \i}
\end{equation}
in order to have more compact expressions. Then
\begin{equation}
     \r_n (\la|\k, \k') = \frac{\La_n (\la|\k')}{\La_0 (\la|\k)}
        = q^{s + \k' - \k} \exp \biggl\{- \int_{{\cal C}_n}
	                   \rd \m \: \re(\m - \la) z_n (\la|\k, \k') \biggr\} \epp
\end{equation}
The eigenvalue ratios appearing in the form-factor series are
recovered by setting $\k' = \k$.

In \cite{DGK13a} we derived expressions for the amplitudes
in the form-factor series of a generating function of the
longitudinal two-point functions and for the amplitudes that
determine the transverse correlation functions $\<\s_1^- \s_{m+1}^+ (t)\>_T$.
Let us recall only the transverse case as an example here. In this
case the relevant amplitudes needed in (\ref{ffseriesnfinite}) are
\begin{equation}
     \frac{\<\Ps_0|B(\x|\k)|\Ps_n\>\<\Ps_n|C(\x|\k)|\Ps_0\>}
     {\La_n (\x|\k) \<\Ps_0|\Ps_0\> \La_0 (\x|\k) \<\Ps_n|\Ps_n\>}
        = \lim_{\k' \rightarrow \k} A_n^{-+} (\x|\k, \k') \epc
\end{equation}
where
\begin{multline} \label{amp}
     A_n^{-+} (\x|\k, \k') = \frac{\overline{G}_+^- (\x) \overline{G}_-^+ (\x)}
                          {(q^{1 + \k' - \k} - q^{- 1 - \k' + \k})
			   (q^{\k' - \k} - q^{- \k' + \k})} \\[1ex] \times
                     \exp \biggl\{
		        \int_{{\cal C}_n} \rd \la \:
			\ln \bigl( \r_n (\la|\k, \k') \bigr)
			\6_\la z_n (\la|\k, \k')
			    \biggr\}\\[1ex] \times
		     \frac{\det_{\rd m_+, {\cal C}_n}
		           \bigl\{ 1 - \widehat{K}_{1-\k'+\k} \bigr\}
		           \det_{\rd m_-, {\cal C}_n}
		           \bigl\{ 1 - \widehat{K}_{1+\k'-\k} \bigr\}}
			  {\det_{\rd m_0, {\cal C}_n}
			   \bigl\{ 1 - \widehat{K} \bigr\}
			   \det_{\rd m, {\cal C}_n}
			   \bigl\{ 1 - \widehat{K} \bigr\}} \epp
\end{multline}

Here, for $\s = \pm$,
\begin{equation}
     \overline{G}_\s^\pm (\x)
        = \lim_{\Re \la \rightarrow \pm \infty} \overline{G}_\s (\la, \x)
\end{equation}
and $\overline{G}_\pm (\la, \x)$ is the solution of the linear
integral equation
\begin{multline} \label{liegbar}
     \overline{G}_\pm (\la, \x) = 
        - \cth(\la - \x)
	+ q^{\k' - \k \mp 1} \r_n^{\pm 1} (\x|\k, \k') \cth(\la - \x + \i \g) \\
	+ \int_{{\cal C}_n} \rd m_\pm (\m) \overline{G}_\pm (\m, \x)
	  K_{\k' - \k \mp 1} (\m - \la)
\end{multline}
with deformed kernel
\begin{equation}\label{kernelK}
     K_\k (\la) = q^{- \k} \cth(\la + \i \g) - q^\k \cth(\la - \i \g) \epp
\end{equation}
Note that $\overline{G}_-^+$ is analytic in $\k' - \k$ and that
$\overline{G}_-^+|_{\k' = \k} = 0$ which implies that the limit
$\k' \rightarrow \k$ exists in (\ref{amp}) \cite{Dugave15}.

The `measures' $\rd m_\eps$, $\eps = -, 0, +$, and $\rd m$ are defined by
\begin{subequations}
\begin{align}
     \rd m_- (\la) &
        = \frac{\rd \la \: \r_n^{-1} (\la|\k, \k')}{2 \p \i (1 + \fa_0 (\la|\k))} \epc \qd
     \rd m_+ (\la)
        = \frac{\rd \la \: \r_n (\la|\k, \k')}{2 \p \i (1 + \fa_n (\la|\k'))} \epc  \label{measures1}\\
     \rd m (\la) & = \frac{\rd \la}{2 \p \i (1 + \fa_0 (\la|\k))} \epc \qd
     \rd m_0 (\la) = \frac{\rd \la}{2 \p \i (1 + \fa_n (\la|\k'))} \epp \label{measures2}
\end{align}
\end{subequations}
The determinants in (\ref{amp}) are Fredholm determinants of integral
operators defined by the respective kernels and measures and by the
integration contours ${\cal C}_n$ (see \cite{DGK13a} for more details).
Note that the representation \eqref{amp} is well-defined in the
sense that for Bethe states with non-vanishing norm $\<\Psi_n|\Psi_n\>
\not=0$ -- which is what we assume to hold and what is always
achievable for generic inhomogeneities -- all factors, and particularly
the Fredholm determinants in the denominator, are finite. 

\subsection{On the summation of the form-factor series}
The form-factor series (\ref{ffseriesnfinite}) is a sum over all
excitations of the quantum transfer matrix. But typically the
matrix elements satisfy `selection rules' connected with
conservations laws that will make many of them disappear. In
case of the XXZ chain the selection rule that has to be obeyed
is related to the pseudo-spin conservation. For the example of
the transverse correlation functions considered above
pseudo-spin conservation implies that we may restrict ourselves
to excitations with $s = 1$. Generically it will be enough to
consider excitations with fixed $s \in {\mathbb Z}$.

The issue to be discussed in this section is the partial
summation of the form-factor series by means of
multi-dimensional residue calculus. For this purpose the
above description of the form factors and eigenvalue ratios
based on equivalence classes of contours is not convenient.
We shall rather deform those contours into reference contours
${\cal C}_{0, s}$ which brings about an explicit dependence
on particle and hole parameters. The contours ${\cal C}_n$
were characterized by the fact that all Bethe roots of a
given state are located inside the contour while all other
zeros of $1 + \fa_n$ are outside. After the deformation some
of the Bethe roots will be outside the new reference contour
${\cal C}_{0, s}$, some zeros of $1 + \fa_n$ which are no
Bethe roots will be inside. We shall call the former particles,
the latter holes and denote their numbers by $n_p$, $n_h$,
respectively.

There is a degree of arbitrariness in the choice of the reference
contour. A choice that might appear natural would be
a contour that contains all Bethe roots and no other zeros
of the auxiliary function pertaining to an eigenvalue of
largest modulus in the pseudo-spin-$s$ sector. Due to
(\ref{cnmonodromy}) this choice implies that the function
$\ln (1 + \fa_n)$ has nontrivial monodromy along such contour
unless $s = 0$. Our experience with the case $s = 1$ in the
low-temperature limit \cite{DGK14a} and with $\D = 0$ (see
below) suggests that we obtain simpler formulae if we slightly
deform the reference contour such as to include $s$ more holes,
if $s$ is positive, or to exclude $- s$ more particles, if
$s$ is negative. Then the equation
\begin{equation} \label{czerosmonodromy}
     \int_{{\cal C}_{0, s}} \frac{\rd \m}{2 \p \i}
        \frac{\fa_n' (\m|\k)}{1 + \fa_n (\m|\k)} =
	n_h - n_p - s = 0
\end{equation}
connects the numbers of particles and holes defined
with respect to ${\cal C}_{0, s}$ with the pseudo-spin.
In the following we shall assume for simplicity that
(\ref{czerosmonodromy}) is satisfied. We emphasize, however,
that this is mostly motivated by our aim to give the
formulae a simpler appearance and that this assumption
is inessential for the argument that will allow us to
partially sum the form-factor series below.

All equations and expressions considered in the previous 
subsection can be rewritten with respect to the reference
contour ${\cal C}_{0,s}$. The auxiliary functions $\fa_n$,
the eigenvalue ratios $\r_n$, and the amplitudes $A_n$ then
become explicit functions of the particle and hole roots.
Instead of equivalence classes of contours ${\cal C}_n$
we then use sets of holes $\{x_j^{(n)}\}_{j=1}^{n_h}$
and particles $\{y_k^{(n)}\}_{k=1}^{n_p}$ to classify
the solutions, meaning that we have altogether three
equivalent parameterizations: by sets of Bethe roots,
by equivalence classes of contours, or by sets of particles
and holes associated with a reference contour ${\cal C}_{0,s}$.

In order to understand this in more detail we shall consider
a nonlinear integral equation in which the driving terms depend
explicitly on the particles and holes. We will use the shorthand
notation 
\begin{equation}
     \th (\la) = - \i \ln \biggl(
                   \frac{\sh(\i \g + \la)}{\sh(\i \g - \la)} \biggr) \epp
\end{equation}
This function is called the bare phase. Starting from
(\ref{auxexp}) it is not difficult to come to the following
characterization of the auxiliary functions $\fa_n$.
We fix a reference contour ${\cal C}_{0, s}$. With respect to
this contour we first define a multi-parametric function
$\fa (\la|\{u\}, \{v\}, \k)$ (not to be confused with the
function defined in (\ref{defaux})) as the solution of the
nonlinear integral equation
\begin{multline} \label{nlieuv}
     \ln \bigl( \fa (\la|\{u\}, \{v\}, \k) \bigr) =
	\i \p s + 2 \i \g \k - \ks \re_N (\la)/T
        + \i \sum_{j=1}^{n_h} \th (\la - u_j) - \i \sum_{k=1}^{n_p} \th (\la - v_k) \\
	- \int_{{\cal C}_{0, s}} \frac{\rd \m}{2 \p \i} \: K(\la - \m)
	  \ln (1 + \fa) (\m|\{u\}, \{v\}, \k) \epp
\end{multline}
Here $\ln(1 + \fa)$ is defined along the contour ${\cal C}_{0, s}$
in a similar way as in (\ref{deflogoneplusa}) 
and is further required to have trivial monodromy along this contour.
The function $\fa (\la|\{u\}, \{v\}, \k)$ depends holomorphically
on two sets of variables $\{u\} = \{u_j\}_{j=1}^{n_h}$ and $\{v\}
= \{v_k\}_{k=1}^{n_p}$, where the $u_j$ take values inside
${\cal C}_{0, s}$ and the $v_k$ take values outside. Then solutions
$\{x\} = \{x_j\}_{j=1}^{n_h}$, $\{y\} = \{y_k\}_{k=1}^{n_p}$ of the
`subsidiary conditions'
\begin{equation} \label{subsi}
     \fa (x_j|\{x\}, \{y\}, \k) = \fa (y_k|\{x\}, \{y\}, \k) = - 1 \epc \qd
        j = 1, \dots, n_h,\ k = 1, \dots, n_p \epc
\end{equation}
define sets of hole and particle roots which are in one-to-one
correspondence with solutions $\{\la_j^{(n)}\}_{j=1}^M$ of the
Bethe Ansatz equations (\ref{baes}) and with the above defined
contours ${\cal C}_n$. Thus, we may label them by the same
superscript `$(n)$' implying that
\begin{equation}
     \fa_n (\la|\k) = \fa (\la|\{x^{(n)}\}, \{y^{(n)}\}, \k) \epp
\end{equation}

The advantage of the `particle-hole formulation' of the excitations
is, as we shall see, that within this formulation all terms in
the form-factor series (\ref{ffseriesnfinite}) can be interpreted
as multi-dimensional residues. Using the properties of the function
$\fa (\la|\{u\}, \{v\}, \k)$ the series can then be turned into a sum
over multiple integrals.

The terms in the form-factor series are composed of eigenvalue
ratios and amplitudes. If we deform the contours into ${\cal C}_{0,s}$,
the eigenvalue ratios take the form
\begin{multline} \label{evratcszero}
     \r_n (\la|\k, \k') = q^{\k' - \k}
	  \biggl[ \prod_{j=1}^{n_h} \frac{\sh(\la - x_j^{(n)})}
	                                 {\sh(\la - x_j^{(n)} + \i \g)} \biggr]
	  \biggl[ \prod_{k=1}^{n_p} \frac{\sh(\la - y_k^{(n)} + \i \g)}
	                                 {\sh(\la - y_k^{(n)})} \biggr] \\ \times
          \exp \biggl\{ - \int_{{\cal C}_{0,s}} \rd \m \: \re(\m - \la)
                                                z_n (\m|\k, \k') \biggr\} \epp
\end{multline}
Note that the function $z_n$ under the integral depends on $\fa_n$ and
is also parameterized by sets of particles and holes. We could proceed
with the different factors appearing in the representation (\ref{amp})
of the amplitudes and make the dependence on the particle and hole
parameters explicit. The exponential term was treated in general in
\cite{DGK14a}. So far the remaining factors were only considered in
the low-temperature limit \cite{DGK13a,DGK14a,DGKS16b}. But the case
of arbitrary temperature is no more difficult. The important point
is that the argument that follows below is based on the fact that these
terms can be parameterized in terms of particles and holes, but does
not depend on the details of the parameterization.

For this reason we refrain here from working out all details but
rather concentrate on the determinant $\det_{\rd m_0, {\cal C}_n}
\bigl\{ 1 - \widehat{K} \bigr\}$ in the denominator. As we
shall see, a Jacobian can be factored out from this term,
whose structure suggests to use multiple-residue calculus
for the summation over the excitations with a fixed number
of particles and holes. This was observed before in the analysis
of the low-temperature limit of the static two-point
functions in the massive regime $\D > 1$ \cite{DGKS16b}
and even earlier in the analysis of the form factor of
the usual transfer matrix for $\D > 1$ in \cite{DGKS15a}. In
Appendix~\ref{app:extractnorm} we show that such a Jacobian
appears in general. We derive the identity
\begin{multline} \label{straightnorm}
     \det_{\rd m_0, {\cal C}_n} \bigl\{ 1 - \widehat{K} \bigr\} = 
          \det_{\rd m_0, {\cal C}_{0,s}} \bigl\{ 1 - \widehat{K} \bigr\}
     \biggl[ \prod_{j=1}^{n_h} \frac{1}{\fa_n' (x_j^{(n)}|\k)} \biggr]
     \biggl[ \prod_{j=1}^{n_p} \frac{1}{\fa_n' (y_j^{(n)}|\k)} \biggr] \\ \times
     \det
	\begin{vmatrix}
	   \6_{u_k} \fa (u_j|\{u\}, \{v\}, \k) & \6_{v_k} \fa (u_j|\{u\}, \{v\}, \k) \\
	   \6_{u_k} \fa (v_j|\{u\}, \{v\}, \k) & \6_{v_k} \fa (v_j|\{u\}, \{v\}, \k) 
	\end{vmatrix}_{\substack{\{u\} = \{x^{(n)}\}\\ \{v\} = \{y^{(n)}\}}} \epp
\end{multline}
Here the products over reciprocals of $\fa_n'$ will be canceled
by corresponding terms originating from the exponential factor in
(\ref{amp}).

The determinant on the right hand side of equation (\ref{straightnorm}) 
is exactly what is needed (see \cite{AiYu83,Range98,DGKS15a,DGKS16b})
to transform a sum over solutions of the subsidiary conditions
(\ref{subsi}) into a multiple-contour integral over `particle
and hole variables' $u_j$ and $v_j$. It may be interpreted as the
Jacobian $\6(\CapitalUv,\CapitalVv)/\6(\uv,\vv)$ of a transformation
${\mathbb C}^{n_h + n_p} \mapsto {\mathbb C}^{n_h + n_p}$,
$(\uv, \vv) \mapsto (\CapitalUv, \CapitalVv)$, where
\begin{subequations}
\label{aasamap}
\begin{align}
     U_j (\uv, \vv) & = 1 + \fa (u_j|\{u\},\{v\},\k) \epc
                        \qd j = 1, \dots, n_h \epc \\[1ex]
     V_k (\uv, \vv) & = 1 + \fa (v_k|\{u\},\{v\},\k) \epc
                        \qd k = 1, \dots, n_p \epp
\end{align}
\end{subequations}
This transformation maps solutions to the subsidiary conditions
(\ref{subsi}) to the origin in ${\mathbb C}^{n_h + n_p}$,
\begin{equation}
     (\xv^{(n)}, \yv^{(n)}) \mapsto (\CapitalUv, \CapitalVv) = (0,0) \epp
\end{equation}
We shall assume that the map is invertible in the neighbourhood of
$(\xv^{(n)}, \yv^{(n)})$, viz.\ that the Jacobian
\begin{equation}
     \Bigl[\frac{\6 (\CapitalUv,\CapitalVv)}{\6 (\uv,\vv)}\Bigr]^{(n)} =
        \frac{\6 (\CapitalUv,\CapitalVv)}{\6 (\uv,\vv)} \biggr|_{(\uv,\vv) = (\xv^{(n)}, \yv^{(n)})}
\end{equation}
is non-vanishing. Let
\begin{equation}
      D_{\eps,\h}^{(n)} =
         \Biggl\{ (\uv, \vv) \in {\mathbb C}^{n_h + n_p}  \; \Bigg|
            \begin{array}{l} 
	         |U_j (\uv, \vv)| < \eps \epc \; |V_k (\uv, \vv)| < \eps \\[.5ex] 
		 \bigl| (\xv^{(n)}, \yv^{(n)}) - (\uv, \vv) \bigr| < \h  
            \end{array} 
         \Biggr\} \epc
\end{equation}
where $\eps$ and $\h$ are sufficiently small so that $D_{\eps, \h}^{(n)}$ is
included in the domain where the map (\ref{aasamap}) is invertible and holomorphic.
Then, for any function $f(\uv,\vv)$ which is holomorphic in all
$u_j$ and $v_k$, we obtain the local residue
\begin{equation} \label{localresi}
     \int_{ {\cal S}_{\eps, \eta}^{(n)} }
        \frac{\rd u^{n_h}}{(2 \p \i)^{n_h}} \frac{\rd v^{n_p}}{(2 \p \i)^{n_p}}
	\frac{f(\uv, \vv)}{\Bigl[\prod_{j=1}^{n_h} U_j (\uv, \vv)\Bigr]
	                   \Bigl[\prod_{k=1}^{n_p} V_k (\uv, \vv)\Bigr]} =
     \frac{f(\xv^{(n)}, \yv^{(n)})}
          {\bigl[\frac{\6 (\CapitalUv,\CapitalVv)}{\6 (\uv,\vv)}\bigr]^{(n)}} \epp
\end{equation}
Here 
\begin{equation}
     {\cal S}_{\eps, \eta}^{(n)}  =
         \Biggl\{ (\uv, \vv) \in {\mathbb C}^{n_h + n_p}  \; \Bigg|
	    \begin{array}{l} 
	         |U_j (\uv, \vv)| = \eps \epc \; |V_k (\uv, \vv)| = \eps \\[.5ex]
		 \bigl| (\xv^{(n)}, \yv^{(n)}) - (\uv, \vv) \bigr| < \h  
            \end{array} 
          \Biggr\} \epp
\end{equation}

If we now make the dependence on the particle and hole parameters
explicit in every term and use (\ref{straightnorm}), the summands in
(\ref{ffseriesnfinite}) take the form
\begin{equation}
     F^{-+} (\{x^{(n)}\}|\{y^{(n)}\}) \Big/
          \Bigl[\frac{\6 (\CapitalUv,\CapitalVv)}{\6 (\uv,\vv)}\Bigr]^{(n)}
\end{equation}
of local multi-dimensional residues, where
\begin{multline} \label{summandform}
     F^{-+} (\{x^{(n)}\}|\{y^{(n)}\}) = \\
        {\cal A}^{-+} (\x|\{x^{(n)}\}|\{y^{(n)}\})
	\r_n^m (0|\k,\k) \r_n^\frac{N}{2} (\tsuz/N|\k,\k)
	\r_n^{-\frac{N}{2}} (-\tsuz/N|\k,\k)
\end{multline}
and where the `amplitude density' ${\cal A}^{-+}
(\x|\{x^{(n)}\}|\{y^{(n)}\})$ is what we obtain when we make
the dependence on the particle and hole parameters explicit in
(\ref{amp}) and extract the Jacobian.

We now replace the particle and hole parameters $\{y^{(n)}\}$
and $\{x^{(n)}\}$ in (\ref{summandform}) by complex variables
$\{u\}$ and $\{v\}$ which we do not require to satisfy the
subsidiary conditions (\ref{subsi}). This means that $\fa_n (\la|\k)$
is replaced by $\fa (\la|\{u\}, \{v\}, \k)$ everywhere, implying
that we obtain 
\begin{equation}
     z (\la|\{u\},\{v\},\k) =
        \frac{\ln (1 + \fa_0) (\la|\k)
	      - \ln (1 + \fa) (\la|\{u\},\{v\},\k)}{2 \pi \i}
\end{equation}
instead of $z_n (\la|\k,\k)$ and
\begin{multline} \label{evratuv}
     \r (\la|\{u\},\{v\}, \k) =
	  \biggl[ \prod_{j=1}^{n_h} \frac{\sh(\la - u_j)}
	                                 {\sh(\la - u_j + \i \g)} \biggr]
	  \biggl[ \prod_{k=1}^{n_p} \frac{\sh(\la - v_k + \i \g)}
	                                 {\sh(\la - v_k)} \biggr] \\ \times
          \exp \biggl\{ - \int_{{\cal C}_{0,s}} \rd \m \: \re(\m - \la)
                                                z (\m|\{u\},\{v\}, \k) \biggr\} \epp
\end{multline}
instead of $\r_n (\la|\k, \k)$. Consequentially, ${\cal A}^{-+}
(\x|\{x^{(n)}\}|\{y^{(n)}\})$ is replaced by a function
${\cal A}^{-+} (\x|\{u\}|\{v\})$ and $F^{-+} (\{x^{(n)}\}|\{y^{(n)}\})$
by a function $F^{-+} (\{u\}|\{v\})$.

Thus, \eqref{localresi} allows us to recast each individual term
in the series \eqref{ffseriesnfinite} -- specialised to the XXZ
chain setting -- as a multi-dimensional local residue integral. 
By the introduction of a suitable function holomorphic in
all $\{u\}$  and $\{v\}$, one may expect a representation
of the sum over all solutions to the subsidiary conditions in
the form of a multi-dimensional residue integral. This idea is made
precise, under certain reasonable hypotheses, in Appendix~\ref{app:sumxxz}.
The resultant multi-dimensional integrations should be preformed over a
skeleton (a distinguished boundary) defined in (\ref{eq:def_skeleton}).
By analogy with one-dimensional residue calculus, one may expect
that the skeleton can be deformed into ${\cal C}_{0,1}^\ell \times
\cbar_{0,1}^{\ell - 1}$, where the contour $\overline{\cal C}_{0,1}$
encloses all particle roots. Taking into account that $n_p + 1 = n_h$
(which follows from (\ref{czerosmonodromy}) since $s = 1$) we then
obtain the following representation for the transverse two-point
functions
\begin{multline} \label{fftxxzfiniten}
     \<\s_1^- \s_{m+1}^+ (t)\>_T =
        \lim_{\substack{N \rightarrow \infty\\ \e \rightarrow 0}}
        \sum_{\ell=1}^\infty \frac{(-1)^\ell \re^{- \i h t}}{\ell! (\ell-1)!}
	\int_{{\cal C}_{0,1}} \frac{\rd u^\ell}{(2 \p \i)^\ell}
	\int_{\overline{\cal C}_{0,1}} \frac{\rd v^{\ell-1}}{(2 \p \i)^{\ell-1}}
	{\cal A}^{-+} (\e|\{u\}|\{v\}) \\[.5ex] \times
	\frac{\r^m (\e|\{u\},\{v\},\k) \, \r^{N/2} (\tsuz/N|\{u\},\{v\},\k) \,
	      \r^{- N/2} (-\tsuz/N|\{u\},\{v\},\k)}
	     {\bigl[\prod_{j=1}^\ell \bigl(1 +
	            \faq \bigl(u_j|\{u\}, \{v\}, \k\bigr)\bigr)\bigr]
	      \bigl[\prod_{k=1}^{\ell-1} \bigl(1 +
	            \fa \bigl(v_k|\{u\}, \{v\}, \k\bigr)\bigr)\bigr]} \epp
\end{multline}
When calculating the integrals over the extended contours
${\cal C}_{0,1}$, $\overline{\cal C}_{0,1}$ we will obtain each
local residue with multiplicity $\ell! (\ell - 1)!$ due to the
symmetry of the functions in the denominator under the integral,
which is why we divided each summand by this factor. Note that
the determination of ${\cal C}_{0,1}$ and $\cbar_{0,1}$ in a way
suitable for numerical calculations may be a subtle issue.

Morally we can understand equation (\ref{fftxxzfiniten}) as follows.
By shrinking the contours ${\cal C}_{0,1}$, $\overline{\cal C}_{0,1}$
one picks up the contributions of all solutions to the subsidiary
conditions.  In principle, one should then also pick up contributions
originating from the poles of $F^{-+} (\{u\}|\{v\})$. Indeed, this function
is not a holomorphic function of the $u_j$ and $v_k$. As can be seen
from (\ref{evratuv}), the factor $\r^{- N/2} (-\tsuz/N|\{u\},\{v\},\k)$
has $N/2$-fold poles at $u_j = - \tsuz/N$, $j = 1, \dots, n_h$, inside
${\cal C}_{0, s}$. However, these poles will be compensated by the
functions $\faq (u_j|\{u\}, \{v\}, \k) = 1/\fa (u_j|\{u\}, \{v\}, \k)$
for $j = 1, \dots, n_h$. In fact these functions as well have poles
of order $N/2$ at $- \tsuz/N$ as can be inferred from (\ref{trotterdecomp}),
(\ref{defen}) and (\ref{nlieuv}). Similarly, the factor
$\r^{- N/2} (- \tsuz/N|\{u\},\{v\},\k)$ has poles of order
$N/2$ at $v_k = \i \g - \tsuz/N$, $k = 1, \dots, n_p$, outside
${\cal C}_{0, s}$ close to where we expect the particles
to be located. These poles will be canceled by the functions
$\fa (v_k|\{u\}, \{v\}, \k)$ as can be seen again from
(\ref{trotterdecomp}), (\ref{defen}) and (\ref{nlieuv}).
Moreover, the amplitude densities ${\cal A}^{-+} (\x|\{u\}|\{v\})$
for $\x = 0$ have simple poles at $u_j = 0$, $j = 1, \dots, n_h$,
and at $v_k = \i \g$, $k = 1, \dots, n_p$, which will be
compensated by $\r_n^m (0|\k,\k)$ if $m > 0$.

After rewriting the form-factor series as a sum over multiple
integrals we may finally take the Trotter limit. For this
purpose we introduce the function
\begin{equation}
     E(\la) = \ln \biggl( \frac{\sh(\la)}{\sh(\la - \i \g)} \biggr)
\end{equation}
and remark that
\begin{multline}
     \lim_{\substack{N \rightarrow \infty\\ \e \rightarrow 0}}
	      \r^m (\e|\{u\},\{v\},\k) \, \r^{N/2} (\tsuz/N|\{u\},\{v\},\k) \,
	      \r^{- N/2} (-\tsuz/N|\{u\},\{v\},\k) \\ =
     \exp \biggl\{ \sum_{j=1}^{n_h} \bigl( m E(u_j) - \tsuz \re(u_j) \bigr) -
                   \sum_{j=1}^{n_p} \bigl( m E(v_j) - \tsuz \re(v_j) \bigr) \\ 
		   - \int_{{\cal C}_{0, s}} \rd \m \: z^{\rm lim} (\m |\{u\},\{v\}, \k)
		     \bigl(m \re (\m) - \tsuz \re' (\m) \bigr) \biggr\} \epc
\end{multline}
which follows from (\ref{evratuv}) and where the superscript `$\rm lim$'
refers to the Trotter limit. Then we end up with the thermal form-factor
series representation
\begin{multline} \label{ffstransverse}
     \<\s_1^- \s_{m+1}^+ (t)\>_T =
        \sum_{n=1}^\infty \frac{(-1)^n}{n! (n-1)!}
	\int_{{\cal C}_{0,1}} \!\! \frac{\rd u^n}{(2 \p \i)^n} \!
	\int_{\overline{\cal C}_{0,1}} \!\! \frac{\rd v^{n-1}}{(2 \p \i)^{n-1}} \\ \times
	\biggl[ \prod_{j=1}^n
	        \frac{\re^{m E(u_j) - \tsuz \re(u_j)}}
	             {1 + \faq^{\sf lim} \bigl(u_j|\{u\}, \{v\}, \k\bigr)} \biggr]
     \biggl[ \prod_{j=1}^{n-1}
	        \frac{\re^{- m E(v_j) + \tsuz \re(v_j)}}
	             {1 + \fa^{\sf lim} \bigl(v_j|\{u\}, \{v\}, \k\bigr)} \biggr] \\ \times
       {\cal A}^{-+}_{\rm lim} (0|\{u\},\{v\}) \,
       \re^{- \i h t - \int_{{\cal C}_{0, 1}} \rd \m \: z^{\rm lim} (\m |\{u\},\{v\}, \k)
		     \bigl(m \re (\m) - \tsuz \re' (\m) \bigr)}
\end{multline}
for the transverse correlation functions of the XXZ chain.
In this formula $\fa^{\sf lim}$ denotes the Trotter limit
of the function $\fa$ which is obtained by replacing $\re_N$
with $\re$ in equation (\ref{nlieuv}). Similarly $z^{\rm lim}$
and ${\cal A}_{\rm lim}^{-+}$ are obtained from $z$ and
${\cal A}^{-+}$ by replacing $\fa$ with $\fa^{\sf lim}$.

It should be clear from our derivation that a similar form
factor series representation can be also derived for the
longitudinal correlation functions. In our derivation we used
the implicit assumption that the reference contour
${\cal C}_{0, 1}$ and the contour $\overline{\cal C}_{0,1}$
can be chosen independently of the excitation. In the
low-temperature limit \cite{DGK13a,DGK14a,DGKS16b} and for
$\D = 0$ (see below) we know that this is possible. The general
case will need further study. In fact, the main difficulty
imposed by the series (\ref{ffstransverse}) is that we
have insufficient knowledge about the general Bethe root
patterns at generic temperature. Moreover, it is not always
clear if one can choose ${\cal C}_{0, 1}$ and $\overline{\cal C}_{0,1}$
of an appropriate size and shape, which might be needed to
perform e.g.\ numerical calculations without technical difficulties.
An appropriate starting point for studying (\ref{ffstransverse})
may be the high-temperature limit, where certain simplifications
are expected to occur. At least the longitudinal dynamical
correlation functions remain nontrivial even at infinite
temperature (see e.g.\ \cite{FaMc98,Niemeijer67} and our
discussion of the XX case below).

\subsection{The XX chain} \label{xxsection}
The previous section shows that the summands in the form-factor
series (\ref{ffseriesnfinite}) can be calculated for the XXZ chain
and that a summation for fixed numbers of particles and holes
can be at least formally achieved by multiple-contour integration.
A crucial question will be how efficient the formulae can be made.
The problem with the general XXZ case is that the patterns of Bethe
roots vary as functions of anisotropy parameter, magnetic field and
temperature and that no general classification is known. The only case
in which we fully understand where the Bethe roots are located at
any temperature is the case of the XX chain. As a further test and
in order to provide more explicit examples, we therefore proceed
with the two-point functions of the XX chain.

By definition the XX Hamiltonian is the XXZ Hamiltonian (\ref{hxxz})
with $\D = 0$ corresponding to $\g = \p/2$ in our parameterization.
For this specific value of $\g$ we have
\begin{equation}
     \ks = - 2 \i J \epc \qd \a = \frac{\i h}{\p} \epp
\end{equation}
Our basic bare functions become
\begin{equation} \label{ekxx}
     \re (\la) = \frac{2}{\sh (2 \la)} \epc \qd K(\la) = 0 \epp
\end{equation}
The fact that the kernel function $K$ is identically zero is the
reason for the severe simplification that occur in this case.

Inserting $\g = \p/2$ into the expression (\ref{auxexp}) for
the auxiliary function $\fa_n$ at finite Trotter number we obtain
\begin{equation} \label{defazero}
     \fa_n (\la|\k) = (-1)^s q^{- 2\k} \prod_{k=1}^{N+1}
                 \frac{\tgh(\la - \n_{2k - 1})}{\tgh(\la - \n_{2k})} \epp
\end{equation}
Unlike in the general XXZ case there is a large degeneracy
among the auxiliary functions here. Any Bethe state corresponds
to a set of roots of one of only two different auxiliary
functions, since $\fa_n (\la|\k) = (- 1)^s \fa_0 (\la|\k)$,
$s = 0, 1 \mod 2$, where $\fa_0$ is the auxiliary function of
the dominant state.

Recalling that $\sum_{k = 1}^{N+1} (\n_{2k} - \n_{2k-1}) =
- \ks/T - 2 \e$ for the Trotter decomposition (\ref{trotterdecomp})
and that $\n_k = \CO (1/N)$ for $k \ne N + 1, N + 2$ we can
calculate the limit
\begin{equation} \label{atrotterxx}
     \lim_{N \rightarrow \infty} \lim_{\e \rightarrow 0} \fa_n (\la|\k)
        = (-1)^s \re^{- \frac{\eps (\la)}{T}} \epc \qd
	  \eps (\la) = h - \frac{4 \i J}{\sh(2 \la)}
\end{equation}
directly from (\ref{defazero}). This result is compatible with the
general consideration of the previous section. Using (\ref{ekxx}) in
(\ref{nlieexc}) we obtain again (\ref{atrotterxx}).
\begin{remark}
In the XX limit the Bethe Ansatz solution of the eigenvalue problem
of the quantum transfer matrix can be analyzed with full rigour. We
have performed such an analysis for the Trotter decomposition
$\n_k = (-1)^{k+1} \ks /(2NT)$, $k = 1, \dots, 2N$, i.e.\ for the
usual temperature case with no dependence of the inhomogeneity
parameters on time or on $\e$. In this case the Bethe roots are
solutions of the equations $\fa_n (\la|\k) + 1 = 0$, and the
following can be shown
\begin{enumerate}
\item
If $N > 2 J/(\p T)$, then all roots of $\fa_n (\la|\k) + 1 = 0$ are
located inside the strip $0 < \Im \la < \p/2$ modulo $\i \p$.
\item
The patterns of roots are point-symmetric about $\i \p/4$, i.e.\ if
$\la$ is a root then $\i \p/2 - \la$ is a root as well.
\item
The functions $\fa_n (\la|\k) + 1$ with $(-1)^s = \pm 1$ have $2N$ roots
each inside the strip $0 < \Im \la < \p/2$. Denote the sets of these
roots by $S_\pm$. A subset of $S_+$ containing $M$ roots with $N - M$
even or a subset of $S_-$ containing $M$ roots with $N - M$ odd
is called a set of Bethe roots. Sets of Bethe roots are in
one-to-one correspondence with eigenvalues of the quantum transfer
matrix. There are altogether $2^{2N}$ such states, called Bethe states.
\item
All eigenvalues corresponding to Bethe states are mutually distinct,
implying that the quantum transfer matrix has a simple spectrum
and that `the Bethe Ansatz is complete'.
\item
The dominant eigenvalue is the eigenvalue determined by the
unique set of Bethe roots $\{\la_j^{(0)}\}_{j=1}^M$ which is
contained in the strip $0 < \Im \la < \p/4$ and for which $M = N$.
\item
In the Trotter limit $\fa_0 (\la) \rightarrow \re^{-\frac{\eps (\la)}{T}}$,
and a pair of roots $\la_F^\pm$ of $\fa_0 (\la|\k) - 1$ is located
on the line $\Im \la = \p/4$,
\begin{equation}
     \la_F^\pm = \frac{\i \p}{4} \pm \2 \arch \Bigl(\frac{4J}{h} \Bigr) \epp
\end{equation}
These roots will be called the Fermi rapidities.
\end{enumerate}
\end{remark}

In the following we will continue to work with the inhomogeneous
model with Trotter decomposition (\ref{trotterdecomp}). In particular,
we will keep $\e$ and the Trotter number finite until the very last
stage of our calculation. We shall assume, however, that $N$ is
large enough such that the properties of our auxiliary functions
are close to those described in the above remark. This means that
we assume that the dominant state has exactly $N + 1$ Bethe roots
located in the strip $- \p/4 < \Im \la < \p/4$ and that the
corresponding auxiliary function $1 + \fa_0$ has no other zeros
in that strip. 

\begin{figure}
\begin{center}
\includegraphics[width=.95\textwidth]{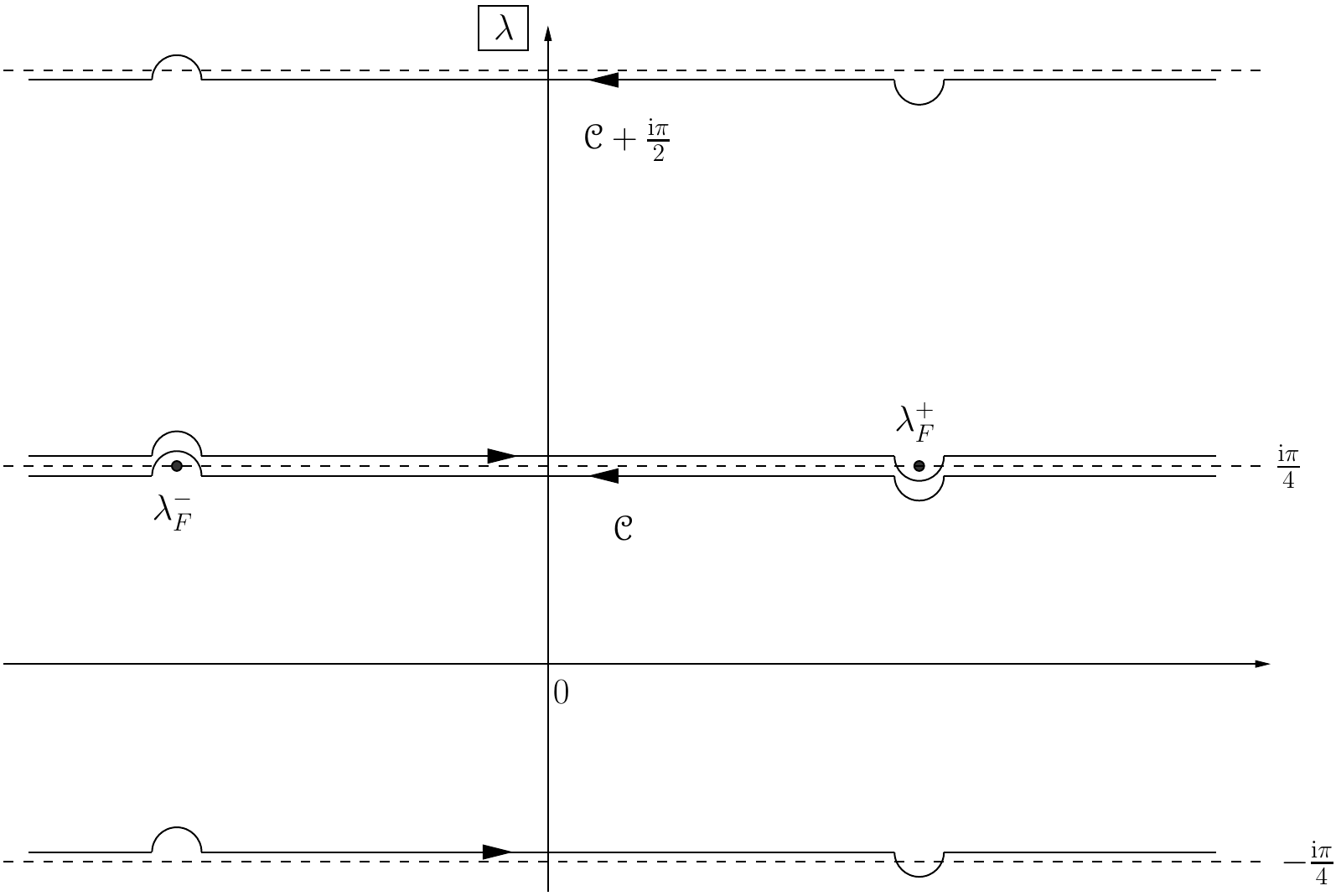}
\caption{\label{fig:xx_transverse_contour} Sketch of the
contour ${\cal C}$ for the XX chain. The Fermi rapidity
$\la_F^-$ is inside ${\cal C}$, $\la_F^+$ is outside ${\cal C}$.}
\end{center}
\end{figure}
Because of the appearance of the Fermi rapidities some care is
necessary when we introduce the canonical contour $\cal C$.
We define it as $(- \infty - \i \p/4 + \i \de, + \infty -
\i \p/4 + \i \de) \cup (+ \infty + \i \p/4 - \i \de,
- \infty + \i \p/4 - \i \de)$, where $\delta > 0$ is small,
but with four small deformations consisting of semicircles of radius
$2\de$, say, which are centered about the points $\la_F^\pm - \i \de$
and $- \la_F^\pm + \i \de$ in such a way that the upper part
of $\cal C$ bypasses $\la_F^+$ from below and $\la_F^-$ from
above, while the lower part of $\cal C$ bypasses $- \la_F^+$ from
above and $- \la_F^-$ from below (see Fig.~\ref{fig:xx_transverse_contour}).
We shall call every zero of $\fa_0 (\la|\k) \pm 1$ inside $\cal C$,
which is not a Bethe root, a hole, while every Bethe root outside
$\cal C$ will be called a particle. The numbers of particles and holes
will be denoted $n_p$ and $n_h$, respectively. Then $M = N + 1
- n_h + n_p$, implying that
\begin{equation} \label{snhnp}
     n_h - n_p = s \epp
\end{equation}

Following the usual reasoning we obtain the following expressions
for the logarithms of the eigenvalues of the quantum transfer
matrix,
\begin{multline} \label{evasxx}
     \ln \bigl( \La_n (\la|\k) \bigr) = - \i \p \k/2 +
        \sum_{k=1}^{n_h} \ln \bigl( \i \tgh (\la_k^h - \la) \bigr) -
        \sum_{k=1}^{n_p} \ln \bigl( \i \tgh (\la_k^p - \la) \bigr) \\
	+ \int_{\cal C} \frac{\rd \m}{\p \i}
	\frac{\ln \bigl(1 + \fa_n (\m|\k)\bigr)}{\sh(2(\m - \la))} \epp
\end{multline}
From this formula we easily deduce the eigenvalue ratios and
their logarithmic derivatives needed in the form-factor series.

\subsubsection{Longitudinal case}
Let us now first consider the example of the longitudinal two-point
functions $\<\s_1^z \s_{m+1}^z (t)\>_T$. For these the operators $X(\x|\k)$
and $Y(\x|\k)$ in (\ref{defamps}) are both equal to $A(\x) - D(\x)
= 2 A(\x) - t(\x|\k)$. Then 
\begin{equation}
     A_0 = \lim_{\substack{N \rightarrow \infty\\ \e \rightarrow 0}}
           \biggl( \frac{\<\Ps_0|(A(\e) - D(\e))|\Ps_0\>}
	                {\<\Ps_0|\Ps_0\> \La_0 (\e|\k)} \biggr)^2 = 4 \fm^2 (T,h)
\end{equation}
is four times the square of the magnetization, and it remains to calculate
\begin{equation}
     A_n = \lim_{\substack{N \rightarrow \infty\\ \e \rightarrow 0}}
           \frac{4 \<\Ps_0|A(\e)|\Ps_n\>\<\Ps_n|A(\e)|\Ps_0\>}
	        {\<\Ps_0|\Ps_0\> \La_n (\e|\k) \<\Ps_n|\Ps_n\> \La_0 (\e|\k)}
\end{equation}
for $n \ne 0$. This can be done by means of Slavnov's scalar product
formula \cite{Slavnov89}. The calculation is rather straightforward but
slightly technical. We show the details in Appendix~\ref{app:detailsxx},
where we arrive at
\begin{equation}
     \frac{\<\Ps_0|A(\x)|\Ps_n\>}{\<\Ps_0|\Ps_0\> \La_n (\x|\k)}
     \frac{\<\Ps_n|A(\x)|\Ps_0\>}{\<\Ps_n|\Ps_n\> \La_0 (\x|\k)} =
        \frac{\re(\x - \la^h)}{\fa_0' (\la^h|\k)}
        \frac{\re(\x - \la^p)}{\fa_0' (\la^p|\k)} \epc
\end{equation}
which is valid for the amplitudes at any finite Trotter number
and $\x$ arbitrary inside the contour. We may set $\x = \e$ and
take the Trotter limit and the limit $\e \rightarrow 0$ which
are determined by equation (\ref{atrotterxx}).

In this very special case all excitations with non-vanishing
amplitudes are parameterized by one particle and one hole rapidity
(see Appendix~\ref{app:detailsxx}). For such excitations $s = 0$
due to (\ref{snhnp}). Using (\ref{evasxx}) we obtain
\begin{equation}
     \r_n (0) = \frac{\tgh(\la^h)}{\tgh(\la^p)} \epc \qd
     \frac{\r_n' (0)}{\r_n (0)} = \re(\la^p) - \re(\la^h) \epp
\end{equation}
Thus, for the longitudinal correlation functions the formal
series (\ref{ffseries}) can be cast into the form
\begin{multline} \label{sumformxxzz}
     \<\s_1^z \s_{m+1}^z (t)\>_T - 4 \fm^2 (T,h) = \\ 4
        \biggl[ \sum_{\la^h}
	\frac{\re (\la^h) \bigl( - \i \tgh(\la^h) \bigr)^m
	      \re^{- \tsuz \re(\la^h)}}{\eps' (\la^h)/T} \biggr]
        \biggl[ \sum_{\la^p}
	\frac{\re (\la^p) \bigl( - \i \tgh(\la^p) \bigr)^{- m}
	      \re^{\tsuz \re(\la^p)}}{\eps' (\la^p)/T} \biggr] \epp
\end{multline}
Here $\exp \{ \tsuz \re(\la) \}$ has an essential singularity at
$\la = 0$ which prevents us from rewriting the series as integrals
and hints that the series are not uniformly convergent in the
excitations.

In order to write the longitudinal two-point functions as an integral
we rather have to use equation (\ref{ffseriesnfinite}). Then
\begin{align} \label{sumformxxzznfinite}
     & \<\s_1^z \s_{m+1}^z (t)\>_T - 4 \fm^2 (T,h) = \\
        & \lim_{\substack{N \rightarrow \infty\\ \e \rightarrow 0}}
        \sum_{\la^h, \la^p}
	\frac{4 \re (\la^h) \re (\la^p)}{\fa_0' (\la^h|\k) \fa_0' (\la^p|\k)}
	\biggl( \frac{\tgh(\la^h)}{\tgh(\la^p)} \biggr)^m
	\biggl( \frac{\tgh(\la^h - \tsuz/N)\tgh(\la^p + \tsuz/N)}
	             {\tgh(\la^h + \tsuz/N)\tgh(\la^p - \tsuz/N)} \biggr)^\frac{N}{2}
		     \epp \notag
\end{align}
The individual terms under the sum have $N/2$-fold poles at
$\la^h = - \tsuz/N$ and at $\la^p = \i \p/2 - \tsuz/N$. Fortunately,
these can be canceled if we choose the auxiliary functions appropriately.
Inserting (\ref{trotterdecomp}) into (\ref{defazero}) we obtain
\begin{equation}
     \fa_0 (\la|\k) = q^{- 2 \k}
	\frac{\tgh(\la - \e)}{\tgh(\la + \e)}
	\biggl[
	\frac{\tgh(\la + \tsuz/N) \tgh(\la - (\tsuz + \ks/T)/N)}
	     {\tgh(\la - \tsuz/N) \tgh(\la + (\tsuz + \ks/T)/N)}
	\biggr]^\frac{N}{2}
\end{equation}
from which we can see that $\fa_0$ has an $N/2$-fold zero at
$- \tsuz/N$ and an $N/2$-fold pole at $\la = \i \p/2 - \tsuz/N$.
Thus, $1 + 1/\fa_0$ has an $N/2$-fold pole at $- \tsuz/N$,
while $1 + \fa_0$ has an $N/2$-fold pole at $\la = \i \p/2 - \tsuz/N$.
Setting $\faq_0 = 1/\fa_0$ it follows for $m > 0$ that
\begin{multline}
     \<\s_1^z \s_{m+1}^z (t) \>_T - 4 \fm^2 (T,h) = \\
        \lim_{\substack{N \rightarrow \infty\\ \e \rightarrow 0}}
        - \biggl[ \int_{\cal C} \frac{\rd \la}{\p \i}
	  \frac{\re (\la) \bigl( - \i \tgh(\la) \bigr)^m}
	       {1 + \faq_0 (\la|\k)}
	\biggl( \frac{\tgh(\la - \tsuz/N)}{\tgh(\la + \tsuz/N)} \biggr)^\frac{N}{2}
	       \biggr] \\[1ex] \times
          \biggl[ \int_{{\cal C} + \frac{\i \p}{2}} \frac{\rd \la}{\p \i}
	  \frac{\re (\la) \bigl( - \i \tgh(\la) \bigr)^{- m}}
	       {1 + \fa_0 (\la|\k)}
	\biggl( \frac{\tgh(\la + \tsuz/N)}{\tgh(\la - \tsuz/N)} \biggr)^\frac{N}{2}
	       \biggr] \epp
\end{multline}
Here the Trotter limit and the limit $\e \rightarrow 0$ required
in (\ref{ffseriesnfinite}) can be taken. Using (\ref{atrotterxx})
we obtain
\begin{multline} \label{szszrapcontour}
     \<\s_1^z \s_{m+1}^z (t) \>_T = 4 \fm^2 (T,h) \\[1ex]
        - \biggl[ \int_{\cal C} \frac{\rd \la}{\p \i}
	  \frac{\re (\la) \bigl( - \i \tgh(\la) \bigr)^m \re^{- \tsuz \re (\la)}}
	       {1 + \re^{\frac{\eps (\la)}{T}}}
	       \biggr]
          \biggl[ \int_{{\cal C} + \frac{\i \p}{2}} \frac{\rd \la}{\p \i}
	  \frac{\re (\la) \bigl( - \i \tgh(\la) \bigr)^{- m} \re^{\tsuz \re(\la)}}
	       {1 + \re^{- \frac{\eps (\la)}{T}}}
	       \biggr] \epp
\end{multline}

This can be transformed into a more familiar form by employing the
$\i \p$-periodicity of the integrand in the second integral and by
turning to momentum variables.
The one-particle momentum is defined as
\begin{equation} \label{defp}
     p (\la) = - \i \ln \bigl( - \i \tgh(\la) \bigr) \epc
\end{equation}
where we understand the logarithm as its principal value, meaning
that we provide cuts in the complex plane from $- \i \p /2$ to zero
modulo $\i \p$. The one-particle momentum is real on the lines
$\Im \la = \pm \p/4$,
\begin{equation} \label{onepmom}
     p(\la) = \begin{cases}
                 - \frac{\p}{2} + 2 \arctg \bigl( \re^{- 2 \Re \la} \bigr)
		 & \text{if $\Im \la = \p/4$} \\
                 - \p \sign(\Re \la) + \frac{\p}{2}
		 - 2 \arctg \bigl( \re^{-2 \Re \la} \bigr)
		 & \text{if $\Im \la = - \p/4$.}
              \end{cases}
\end{equation}
Hence the assignment $\la \mapsto p$ maps
\begin{subequations}
\begin{align}
     & (- \infty - \i \p/4, + \infty - \i \p/4)
          \longmapsto [- \p, -\p/2] \cup [\p/2, \p] \epc \\[1ex]
     & (+ \infty + \i \p/4, - \infty + \i \p/4)
          \longmapsto [-\p/2, \p/2] \epp
\end{align}
\end{subequations}
Then, since the regularizations of the contour play no role in
(\ref{szszrapcontour}),
\begin{equation} \label{xxdyntfine}
     \<\s_1^z \s_{m+1}^z (t)\>_T = 4 \fm^2 (T,h)
        + \biggl[ \int_{- \p}^\p \frac{\rd p}{\p}
	          \frac{\re^{\i (m p - t \e(p))}}
		       {1 + \re^{\e(p)/T}} \biggr]
          \biggl[ \int_{- \p}^\p \frac{\rd p}{\p}
	          \frac{\re^{- \i (m p - t \e(p))}}
		       {1 + \re^{- \e(p)/T}} \biggr] \epc
\end{equation}
where we have introduced the energy function in momentum
variables
\begin{equation}
     \e(p) = h - 4J \cos(p) \epp
\end{equation}
For the sake of completeness let us also recall the expression
\begin{equation}
     \fm(T,h) = \int_{- \p}^\p \frac{\rd p}{4 \p} \tgh \biggl(\frac{\e(p)}{2T}\biggr)
\end{equation}
for the magnetization as a function of temperature and magnetic
field here.

Equation (\ref{xxdyntfine}) is the final result for the longitudinal
finite temperature dynamical two-point correlation function of
the XX chain. Note that this beautifully simple formula is different
from those given in the classical papers \cite{Niemeijer67,KHS70}
(where less natural parameterizations were used), but can be also
obtained within an approach based on mapping the XX model to free
spinless Fermions.

In our derivation of the form-factor series (\ref{ffseriesnfinite})
we required invertibility of the inhomogeneous shift operators. 
This means for the example at hand that we have implicitly
assumed that $t \ne 0$. Nevertheless, the known static case
(see e.g.\ \cite{GoSe05}) is reproduced from (\ref{xxdyntfine})
for $t \rightarrow 0$ using that $(1 + \re^{- \e(p)/T})^{-1} =
1 - (1 + \re^{\e(p)/T})^{-1}$ and that $\int_{-\p}^\p \frac{\rd p}{2 \p} \:
\re^{- \i m p} = \de_{m,0}$. Then
\begin{equation} \label{stattfine}
     \<\s_1^z \s_{m+1}^z\>_T = 4 \fm^2 (T,h) + \de_{m,0} \bigl(2 - 4 \fm (T,h)\bigr)
        - \biggl| \int_{- \p}^\p \frac{\rd p}{\p}
	          \frac{\re^{\i m p}}
		       {1 + \re^{\e(p)/T}} \biggr|^2 \epp
\end{equation}
The known high-temperature limit \cite{Niemeijer67} follows easily
as well if we set $1/T = 0$, implying that
\begin{equation} \label{hightdyn}
     \<\s_1^z \s_{m+1}^z (t)\>_\infty = J_m^2 (4Jt) \epc
\end{equation}
where $J_m$, $m \in {\mathbb N}$, is a Bessel function.

Note that (\ref{xxdyntfine}), even holds if $t = 0$, $m = 0$,
although we assumed $m > 0$ in the derivation. Similarly,
(\ref{stattfine}) and (\ref{hightdyn}) remain valid for $m = 0$.
It seems that the validity in these limiting cases is assured
by analytic continuation in $m$ and $t$.

\subsubsection{Transverse case}
While our study of the longitudinal case in the previous section
basically showed that our form-factor formalism works and reproduces
the known result, it brings about something new when we move on to
the transverse case.

We shall consider the correlation function
$\<\s_1^- \s_{m+1}^+ (t)\>_T$. For this correlation function the
operators $X(\x|\k)$ and $Y(\x|\k)$ in (\ref{defamps}) are equal
to $B(\x)$ and $C(\x)$, respectively. In Appendix~\ref{app:detailsxx}
we calculate the finite Trotter number amplitudes
\begin{equation} \label{defatrans}
     A_n^{-+} (\x) =
           \frac{\<\Ps_0|B(\x)|\Ps_n\>\<\Ps_n|C(\x)|\Ps_0\>}
	        {\<\Ps_0|\Ps_0\> \La_n (\x|\k) \<\Ps_n|\Ps_n\> \La_0 (\x|\k)}
\end{equation}
for small finite $\e$. They are non-zero only for (pseudo-) spin $s=1$
excitations with corresponding auxiliary functions $\fa_n = - \fa_0$ and
are parameterized by sets of hole rapidities $\{\la_j^h\}_{j=1}^{n_h}$
and particle rapidities $\{\la_j^p\}_{j=1}^{n_p}$. For our choice of
contour the numbers of particle and hole rapidities are related by
(\ref{snhnp}). Hence, $n_h = n_p + 1$, and we write $n = n_h$ for
short. We further introduce the functions
\begin{align} \label{defz}
     & z(\la) = \frac{1}{2 \p \i}
                \ln \biggl( \frac{1 + \fa_0 (\la|\k)}{1 + \fa_n (\la|\k)} \biggr)
		\epc \\[1ex] \label{defphi}
     & \PH (x) = \frac{\re(x)}{2} \times
                 \exp \biggl\{ 2 \int_{\cal C} \rd \m \cth (x - \m) z(\m) \biggr\}
                 \epc \\[1ex] \label{defd}
     & {\cal D} \bigl(\{x_j\}_{j=1}^{n_h}, \{y_k\}_{k=1}^{n_p}\bigr) =
        \frac{\bigl[ \prod_{1 \le j < k \le n_h} \sh^2 (x_j - x_k) \bigr]
              \bigl[ \prod_{1 \le j < k \le n_p} \sh^2 (y_j - y_k) \bigr]}
             {\prod_{j=1}^{n_h} \prod_{k=1}^{n_p} \sh^2 (x_j - y_k)}
\end{align}
and the `prefactors'
\begin{subequations}
\begin{align} \label{defprefactorxx}
     & {\cal A} =
        \exp \biggl\{ 2 \int_{\cal C} \rd \m \cth (2 \m) z(\m) -
	             \int_{{\cal C}' \subset {\cal C}} \rd \la
		     \int_{\cal C} \rd \m \cth' (\la - \m) z(\la) z(\m) \biggr\} \epc \\
     & {\cal A} (m,t) = {\cal A} \times
        \exp \biggl\{ - \int_{\cal C} \rd \m \: z(\m)
	                \bigl[ m \re(\m) - \tsuz \re' (\m) \bigr] \biggr\}
			\label{defamt} \epc
\end{align}
\end{subequations}
which depend parametrically on temperature and magnetic field as well.
The contour ${\cal C}'$ in (\ref{defprefactorxx}) is tightly enclosed
by $\cal C$.

Using this notation all amplitudes can be written as
\begin{equation} \label{ampspmxx}
     A_n^{-+} (0) = {\cal A} \times
        \biggl[ \prod_{\la^h \in {\cal H}} \frac{2 \PH (\la^h)}{\fa_n' (\la^h|\k)} \biggr]
	\biggl[ \prod_{\la^p \in {\cal P}} \frac{2 \PH (\la^p)^{-1}}
                                       {\fa_n' (\la^p|\k)} \biggr]
        {\cal D} ({\cal H}, {\cal P}) \epc
\end{equation}		                                
where ${\cal H} = \{\la_j^h\}_{j=1}^n$ and ${\cal P} =
\{\la_k^p\}_{k=1}^{n-1}$ are sets of hole and particle rapidities,
i.e.\ sets of zeros of $1 + \fa_n$ located inside ${\cal C}$
or ${\cal C} + \i \p/2$, respectively. For the eigenvalue ratios
equation (\ref{evasxx}) implies that
\begin{equation} \label{evspmxx}
     \r_n (\la) = \frac{\prod_{\la^h \in {\cal H}} \i \tgh(\la^h - \la)}
                       {\prod_{\la^p \in {\cal P}} \i \tgh(\la^p - \la)}
                  \exp \biggl\{ - \int_{\cal C} \rd \m \: \re (\m - \la) z(\m) \biggr\} \epp
\end{equation}

Inserting (\ref{ampspmxx}) and (\ref{evspmxx}) into (\ref{ffseriesnfinite})
we obtain
\begin{multline} \label{pmxxsumnfinite}
     \<\s_1^- \s_{m+1}^+ (t)\>_T = \re^{- \i h t} \\ \times
        \lim_{\substack{N \rightarrow \infty \\ \e \rightarrow 0}} {\cal A} (m, t)
	\sum_{\cal P, H} \biggl[ \prod_{x \in \cal H}
	   \frac{2 \PH (x)}{\fa_n' (x|\k)} \: \bigl(\i \tgh (x) \bigr)^m
	   \biggl( \frac{\tgh ( x - \tsuz/N)}
	                {\tgh ( x + \tsuz/N)} \biggr)^{\frac N2} \biggr] \\ \times
	   \biggl[ \prod_{x \in \cal P}
	   \frac{2 \PH (x)^{-1}}{\fa_n' (x|\k)} \: \bigl(\i \tgh (x) \bigr)^{- m}
	   \biggl( \frac{\tgh ( x - \tsuz/N)}
	                {\tgh ( x + \tsuz/N)} \biggr)^{- \frac N2} \biggr]
        {\cal D} ({\cal H}, {\cal P}) \epc
\end{multline}
where the sum is over all sets of particles and holes. This sum
can be easily transformed into a sum over multiple integrals. The
discussion about the singularities of the integrands parallels the
discussion below equation (\ref{sumformxxzznfinite}). In particular,
we shall assume that $m > 0$. Then we rewrite the sum in
(\ref{pmxxsumnfinite}) as a sum over multiple integrals, introduce
the one-particle energy (\ref{atrotterxx}) and momentum function
(\ref{onepmom}) and finally perform the limits $N \rightarrow \infty$
and $\e \rightarrow 0$. We arrive at
\begin{multline} \label{ffseriesxxtrans}
     \<\s_1^- \s_{m+1}^+ (t)\>_T = (-1)^m {\cal A} (m,t)
        \sum_{n=1}^\infty \frac{(-1)^n}{n! (n-1)!}
           \int_{\cal C} \prod_{r=1}^n \frac{\rd x_r}{\p \i}
	      \frac{\PH_- (x_r) \re^{\i (m p(x_r) - t \eps (x_r))}}
	           {1 - \re^\frac{\eps (x_r)}{T}} \\ \times
           \int_{\overline{\cal C}} \prod_{s=1}^{n-1} \frac{\rd y_s}{\p \i}
	      \frac{\re^{- \i (m p(y_s) - t \eps (y_s))}}
	           { \PH_- (y_s) \bigl[1 - \re^{-\frac{\eps (y_s)}{T}}\bigr]} \:
	      {\cal D} \bigl( \{x_r\}_{r=1}^n, \{y_s\}_{s=1}^{n-1} \bigr) \epp
\end{multline}
By $\PH_-$ we mean the boundary values of $\PH$ from
inside the contour. $\overline{\cal C}$ is the particle contour
which can be chosen as $\overline{\cal C} = {\cal C} + \i \p/2$.
In this case the regularization introduced above is important.
By ${\cal A} (m,t)$ and $\PH_- (x)$ we now mean the functions
obtained from (\ref{defamt}) and (\ref{defphi}) in the Trotter
limit, i.e.\ by replacing $\fa_0 (\la)$ by $\re^{- \eps(\la)/T}$.
Equation (\ref{ffseriesxxtrans}) provides a novel form-factor
series for the transverse two-point functions of the XX chain.
A detailed analysis of this series will be presented in a separate
work.


\section{Conclusions}
We have devised a thermal form-factor approach to the
dynamical correlation functions of fundamental integrable
lattice models at finite temperature. This approach provides
thermal form-factor series expansions of the two-point
correlation functions of these models. The summands in the
series are determined by ratios of eigenvalues of the
quantum transfer matrix and by amplitudes, which are products
of two thermal form factors. For finite Trotter number both,
the eigenvalue ratios and the amplitudes, depend on time
in the dynamical case. But at least for the XXZ chain this
time dependence vanishes in the Trotter limit in which
eigenvalue ratios and amplitudes are given by the well-known
expressions studied in \cite{DGK13a,DGK14a,DGKS16b}.

Hence, for the XXZ chain, the remaining question is how
to evaluate the series. With the simplest possible example,
namely the two-point functions of the XX chain, we have shown
that the summation can be efficiently performed. We have
reproduced the existing results for the longitudinal case
and have derived a novel form-factor series in the transverse
case that will be the starting point for further studies.
In the general XXZ case we have suggested how the sums
in every sector of fixed particle and hole numbers can
be rewritten as multiple-contour integrals. Our formula
will remain a conjecture until we have gained deeper insight
into the concrete construction of the integration contours.

In future work we plan to proceed with the general XXZ
case. In the most generic situation of arbitrary times
and distances at arbitrary temperatures we expect that
some computer effort will remain. For small and large
temperature we hope to obtain explicit results for the
long-time and large-distance asymptotics. An important goal
of our future work will be to develop a physical intuition
for the behaviour of thermal correlation functions,
particularly for long times and large distances.\\[1.ex]
{\bf Acknowledgments.} FG, MK and AK acknowledge
financial support by the DFG in the framework of the
research group FOR 2316 and through grant number Go 825/9-1.
FG wishes to thank the ENS de Lyon and Shizuoka University
for hospitality. KKK is supported by CNRS. JS is grateful
for support by a JSPS Grant-in-Aid for Scientific Research (C)
No.\ 15K05208.

\clearpage
{\appendix
\Appendix{A proof of the inversion formulae for the quantum transfer
matrix} \label{app:invqtm} \noindent
In this appendix we provide a proof of equation (\ref{soqip}). Without
restriction of generality we may replace $N$ by $N - 1$ in the
definition of the staggered monodromy matrix (\ref{stagmon}). For
every odd $j \in \{1, \dots, 2N\}$ we introduce cyclically reordered
staggered monodromy matrices
\begin{subequations}
\begin{align}
     & T_{a; j, \dots, 2N, 1, \dots, j - 1} (\la|\a) =
        \notag \\ & \mspace{18.mu}
        R_{j-1, a}^{t_1} (\n_{j-1}, \la) \dots R_{a, 1} (\la, \n_1)
	\Th_a (\a) R_{2N, a}^{t_1} (\n_{2N}, \la) \dots
        R_{a, j} (\la, \n_j) \epc \\[1ex]
     & T_{a; j-1, \dots, 2N, 1, \dots, j - 2} (\la|\a) =
        \notag \\ & \mspace{18.mu}
        R_{a, j - 2} (\la, \n_{j-2})
	\dots R_{a, 1} (\la, \n_1)
	\Th_a (\a) R_{2N, a}^{t_1} (\n_{2N}, \la) \dots
        R_{j-1, a}^{t_1} (\n_{j-1}, \la) \epp
\end{align}
\end{subequations}
Then $T_a (\la|\a) = T_{a; 1, \dots, 2N}$.

\noindent \emph{Step 1.} Using (\ref{reg}) we obtain
\begin{align} \label{step1}
     & \tr_a \bigl\{ x_a T_{a ; j, \dots, 2N, 1, \dots, j - 1} (\n_j|\a) \bigr\}
        \notag \\
     & \mspace{18.mu} = x_j
        R_{j-1, j}^{t_1} (\n_{j-1}, \n_j) \dots R_{j, 1} (\n_j, \n_1)
	\Th_j (\a) R_{2N, j}^{t_1} (\n_{2N}, \n_j) \dots
        R_{j+1, j}^{t_1} (\n_{j+1}, \n_j)
        \notag \\
     & \mspace{18.mu} = x_j t(\n_j|\a) \epp
\end{align}
If $j = 1$, (\ref{step1}) reads
\begin{equation}
     \tr_a \bigl\{ x_a T_a (\n_1|\a) \bigr\} = x_1 t(\n_1|\a)
\end{equation}
which proves (\ref{soqip}) for $j = 1$.

\noindent \emph{Step 2.} For $j$ odd and $j > 1$ we have
\begin{align} \label{stepping}
     & t(\n_{j-1}|\a) = 
       \tr_a \bigl\{ R_{a,j-2} (\la, \n_{j-2}) \dots R_{a,1} (\la,\n_1)
                     \Th_a (\a) \notag \\ & \mspace{180.mu} \times
		     R_{2N,a}^{t_1} (\n_{2N},\la) \dots
		     R_{j-1,a}^{t_1} (\n_{j-1},\la) \bigr\}\Bigr|_{\la = \n_{j-1}}
		     \notag \\[1ex] & \mspace{36.mu} =
       \tr_a \bigl\{ R_{j-1,a} (\n_{j-1},\la) R_{a,j}^{t_1} (\la,\n_j)
                     \dots R_{2N,a} (\n_{2N},\la)
                     \Th_a^t (\a) \notag \\ & \mspace{180.mu} \times
		     R_{a,1}^{t_1} (\la,\n_1) \dots
		     R_{a,j-2}^{t_1} (\la,\n_{j-2}) \bigr\}\Bigr|_{\la = \n_{j-1}}
		     \notag \\ & \mspace{36.mu} =
       R_{j-1, j}^{t_1} (\n_{j-1}, \n_j) \dots R_{2N,j-1} (\n_{2N}, \n_{j-1})
                     \Th_{j-1}^t (\a) \notag \\[1ex] & \mspace{180.mu} \times
		     R_{j-1,1}^{t_1} (\n_{j-1},\n_1) \dots
		     R_{j-1,j-2}^{t_1} (\n_{j-1},\n_{j-2}) \epp
\end{align}
Here we have used the cyclicity of the trace in the first
equation, (\ref{sym}) in the second equation and (\ref{reg})
in the third equation. The invariance equation
(\ref{groupinvariancer}) implies that
\begin{equation} \label{invtransp}
     \Th_{j-1}^t (\a) R_{j-1,a}^{t_1} (\n_{j-1},\la) \Th_a (\a) =
	\Th_a (\a) R_{j-1,a}^{t_1} (\n_{j-1},\la) \Th_{j-1}^t (\a) \epp
\end{equation}
Using (\ref{stepping}), (\ref{invtransp}), (\ref{ybe}), (\ref{uni})
and (\ref{ybet1}) it is easy to see that
\begin{equation}
     t(\n_{j-1}|\a) T_{a; j, \dots, 2N,1, \dots, j-1} (\la|\a) =
        T_{a; j-1, \dots, 2N,1, \dots, j-2} (\la|\a) t(\n_{j-1}|\a) \epp
\end{equation}
Similarly using the representation of $t(\n_{j-2}|\a)$ that can
be read off from (\ref{step1}) with $x = \id$ as well as (\ref{ybe}),
(\ref{uni}), (\ref{groupinvariancer}) and (\ref{ybet1}) we obtain
\begin{equation}
     T_{a; j-1, \dots, 2N,1, \dots, j-2} (\la|\a) t(\n_{j-2}|\a)
        = t(\n_{j-2}|\a) T_{a; j-2, \dots, 2N,1, \dots, j-3} (\la|\a) \epp
\end{equation}

The last two equations can be combined to
\begin{multline}
     T_{a; j, \dots, 2N,1, \dots, j-1} (\la|\a) \\ =
	t(\n_{j-2}|\a) t^{-1} (\n_{j-1}|\a)
        T_{a; j-2, \dots, 2N,1, \dots, j-3} (\la|\a)
	t(\n_{j-1}|\a) t^{-1} (\n_{j-2}|\a) \epp
\end{multline}
It follows by iteration that
\begin{multline}
     T_{a; j, \dots, 2N,1, \dots, j-1} (\la|\a) \\ =
        \biggl[\prod_{k=1}^{(j-1)/2} t(\n_{2k-1}|\a)
	       t^{-1} (\n_{2k}|\a) \biggr] T_a (\la|\a)
	\biggl[\prod_{k=1}^{(j-1)/2} t(\n_{2k}|\a)
	       t^{-1} (\n_{2k-1}|\a) \biggr] \epp
\end{multline}
Inserting this into (\ref{step1}) and replacing $N$
by $N + 1$ we have established (\ref{soqip}).

\Appendix{Invertibility of the inhomogeneous shift
operators for the XXZ chain} \label{app:invshift} \noindent
In the previous appendix we have seen that for odd $j$ the
operators
\begin{subequations}
\begin{align} \label{definhomshift}
     t(\n_j|\a) & =
        R_{j-1, j}^{t_1} (\n_{j-1}, \n_j) \dots R_{j, 1} (\n_j, \n_1)
	\Th_j (\a) R_{2N, j}^{t_1} (\n_{2N}, \n_j) \dots
	\notag \\ & \mspace{324.mu}
	\dots R_{j+1, j}^{t_1} (\n_{j+1}, \n_j) \epc \\[1ex]
     t(\n_{j-1}|\a) & = 
       R_{j-1, j}^{t_1} (\n_{j-1}, \n_j) \dots R_{2N,j-1} (\n_{2N}, \n_{j-1})
       \Th_{j-1}^t (\a) \notag \\[1ex] & \mspace{144.mu} \times
       R_{j-1,1}^{t_1} (\n_{j-1},\n_1) \dots
       R_{j-1,j-2}^{t_1} (\n_{j-1},\n_{j-2})
\end{align}
\end{subequations}
shift the monodromy matrix indices cyclically if applied
from the left or right, respectively. In order to
establish sufficient criteria for the invertibility of
the inhomogeneous shift operators for the XXZ chain we
calculate their determinants using the specific form of
$\Th (\a)$ and of the $R$-matrix (\ref{rxxz}).
We will employ the formula $\det( \id_m \otimes A )
= \bigl( \det (A) \bigr)^m$, valid for $A \in
\End ({\mathbb C}^n)$ if $\id_m$ is the identity in
$\End ({\mathbb C}^m)$.

Taking the determinant in (\ref{definhomshift}) we obtain
\begin{multline}
     \det \bigl(t(\n_j|\a)\bigr) = \det \bigl(\Th_j (\a)\bigr)
        \biggl[ \prod_{\substack{k = 1\\k \ne (j+1)/2}}^N
	        \det \bigl( R_{j, 2k - 1} (\n_j, \n_{2k - 1}) \bigr) \biggr]
		\\[-1ex] \times
        \biggl[ \prod_{k = 1}^N
	        \det \bigl( R^{t_1}_{2k, j} (\n_{2k}, \n_j) \bigr) \biggr] \epp
\end{multline}
The site indices can be shifted by means of permutation
operators $P_{jk} = {e_j}_\a^\be {e_k}_\be^\a$, where
$e_\a^\be$ are the canonical matrix units having a single
non-zero matrix element one in the $\a$th row and $\be$th
column. The $P_{jk}$ are invertible, since $P_{jk}^2 = \id$
which also implies that $\det^2 (P_{jk}) = 1$. It follows that
\begin{equation}
     \det (P_{jk}) = \det (P_{2N-1,2N})
        = \bigl( \det(P) \bigr)^{2^{2N-2}} = 1 \epc
\end{equation}
since $\det P = - 1$ and $N$ is even. Hence,
\begin{multline}
     \det \bigl(t(\n_j|\a)\bigr) = \bigl( \det \bigl(\Th (\a)\bigr)\bigr)^{2^{2N-1}}
        \biggl[ \prod_{\substack{k = 1\\k \ne (j+1)/2}}^N
	        \det \bigl( R (\n_j, \n_{2k - 1}) \bigr) \biggr]^{2^{2N-2}}
		\\[-1ex] \times
        \biggl[ \prod_{k = 1}^N
	        \det \bigl( R^{t_1} (\n_{2k}, \n_j) \bigr) \biggr]^{2^{2N-2}} \epp
\end{multline}
Now
\begin{align}
     & \det \bigl( \Th (\a) \bigr) = 1 \epc \qd
     \det \bigl(R(\la,\m)\bigr)
        = \frac{\sh(\la - \m + \i \g)}{\sh(\la - \m - \i \g)} \epc \\
     & \det \bigl(R^{t_1} (\la,\m)\bigr)
        = \frac{\sh^3 (\la - \m) \sh(\la - \m - 2 \i \g)}
	       {\sh^4 (\la - \m - \i \g)} \epp
\end{align}
Thus,
\begin{multline}
     \det \bigl(t(\n_j|\a)\bigr) = \\
        \biggl[ \prod_{k = 1}^N
	        \frac{\sh(\n_j - \n_{2k-1} + \i \g)}
		     {\sh(\n_j - \n_{2k-1} - \i \g)}
		\frac{\sh^3 (\n_{2k} - \n_j) \sh(\n_{2k} - \n_j - 2 \i \g)}
		     {\sh^4 (\n_{2k} - \n_j - \i \g)}
	        \biggr]^{2^{2N-2}} \mspace{-36.mu} \epp
\end{multline}
Similarly
\begin{multline}
     \det \bigl(t(\n_{j-1}|\a)\bigr) = \\
        \biggl[ \prod_{k = 1}^N
	        \frac{\sh(\n_{2k} - \n_{j-1} + \i \g)}
		     {\sh(\n_{2k} - \n_{j-1} - \i \g)}
		\frac{\sh^3 (\n_{j-1} - \n_{2k-1}) \sh(\n_{j-1} - \n_{2k-1} - 2 \i \g)}
		     {\sh^4 (\n_{j-1} - \n_{2k-1} - \i \g)}
	        \biggr]^{2^{2N-2}} \mspace{-36.mu} \epp
\end{multline}

From the latter two equations we infer that
\begin{subequations}
\begin{align}
     & |\n_k| < \g/2 \epc \qd k = 1, \dots , 2N \epc \\
     & \n_{2j-1} \ne \n_{2k} \epc \qd j, k = 1, \dots, N \epc
\end{align}
\end{subequations}
is a set of sufficient conditions for all inhomogeneous shift
operators connected with the inhomogeneous quantum transfer matrix
of the XXZ model to be invertible.

\Appendix{Deforming the contour in a norm determinant}
\label{app:extractnorm}
In order to deform the contour in $\det_{\rd m_0, {\cal C}_n}
\bigl\{ 1 - \widehat{K} \bigr\}$ to ${\cal C}_{0,s}$ we consider the
action of $1 - \hat K$ on a function $f$ that is holomorphic on
${\cal C}_n - {\cal C}_{0, s}$,
\begin{multline}
     (1 - \hat K) f (x) =
        f(x) - \int_{{\cal C}_n} \frac{\rd y}{2 \p \i}\:
	                         \frac{K(x - y) f(y)}{1 + \fa_n(y|\k)} \\
	= f(x) + \sum_{k=1}^{n_h} \frac{K(x - x_k) f(x_k)}{\fa_n' (x_k|\k)}
	       - \sum_{k=1}^{n_p} \frac{K(x - y_k) f(y_k)}{\fa_n' (y_k|\k)}
               - \int_{{\cal C}_{0,s}} \frac{\rd y}{2 \p \i}\:
	         \frac{K(x - y) f(y)}{1 + \fa_n(y|\k)} \epp
\end{multline}
The latter equation allows us to interpret $1 - \hat K$ as a
linear operator acting on functions supported on the set
${\cal C}_{0,s} \cup \{x_j\}_{j=1}^{n_h} \cup \{y_j\}_{j=1}^{n_p}$.
Then the Fredholm determinant can be interpreted as
\begin{align}
     & \det_{\rd m_0, {\cal C}_n} \bigl\{ 1 - \widehat{K} \bigr\} = \\ &
     \det_{{\cal C}_{0,s} \cup \{x_j\}_{j=1}^{n_h} \cup \{y_j\}_{j=1}^{n_p}}
        \begin{vmatrix}
	   \de (x - y) \rd y - \frac{K(x - y) \rd y}{2 \p \i (1 + \fa_n(y|\k))} &
	   \frac{K(x - x_k)}{\fa_n' (x_k|\k)} &
	   - \frac{K(x - y_k)}{\fa_n' (y_k|\k)} \\[1ex]
	   - \frac{K(x_j - y) \rd y}{2 \p \i (1 + \fa_n(y|\k))} &
	   \de_{jk} + \frac{K(x_j - x_k)}{\fa_n' (x_k|\k)} &
	   - \frac{K(x_j - y_k)}{\fa_n' (y_k|\k)} \\[1ex]
	   - \frac{K(y_j - y) \rd y}{2 \p \i (1 + \fa_n(y|\k))} &
	   \frac{K(y_j - x_k)}{\fa_n' (x_k|\k)} &
	   \de_{jk} - \frac{K(y_j - y_k)}{\fa_n' (y_k|\k)}
	\end{vmatrix} \epp \notag
\end{align}
Here we can apply the formula
\begin{equation} \label{extractdet}
     \det \begin{pmatrix} A & B \\ C & D \end{pmatrix}
        = \det (A) \det \begin{pmatrix}
	                   \id & A^{-1} B \\ 0 & D - C A^{-1} B
			\end{pmatrix} \epc
\end{equation}
valid if $A$ is an invertible square matrix,
to extract the determinant of the integral operator in the upper
left corner. The remaining determinant of a finite matrix takes
a simple form in terms of the resolvent kernel defined by
\begin{equation} \label{resolventkernel}
     R(x,y) = K(x - y) +
              \int_{{\cal C}_{0,s}} \frac{\rd z}{2 \p \i}
	                            \frac{K(x - z) R(z,y)}{1 + \fa_n (z|\k)} \epp
\end{equation}
Using (\ref{extractdet}) and (\ref{resolventkernel}) we end up with
\begin{equation} \label{detresolventform}
     \det_{\rd m_0, {\cal C}_n} \bigl\{ 1 - \widehat{K} \bigr\} =
        \det_{\rd m_0, {\cal C}_{0,s}} \bigl\{ 1 - \widehat{K} \bigr\}
	\det
	\begin{vmatrix}
	     \de_{jk} + \frac{R(x_j, x_k)}{\fa_n'(x_k|\k)} &
	     - \frac{R(x_j, y_k)}{\fa_n'(y_k|\k)} \\[1ex]
	     \frac{R(y_j, x_k)}{\fa_n'(x_k|\k)} &
	     \de_{jk} - \frac{R(y_j, y_k)}{\fa_n'(y_k|\k)}
	\end{vmatrix} \epp
\end{equation}

Next we observe that
\begin{equation} \label{resaux}
     R(\la, x_k) = - \frac{\6_{x_k} \fa_n (\la|\k)}{\fa_n (\la|\k)} \epc \qd
     R(\la, y_k) = \frac{\6_{y_k} \fa_n (\la|\k)}{\fa_n (\la|\k)} \epc
\end{equation}
which can be inferred by taking the derivatives with respect to $u_k$
and $v_k$ in the nonlinear integral equation
\begin{multline} \label{nliefaqnczero}
     \ln \bigl( \faq (\la|\{u\}, \{v\}, \k) \bigr) =
        - 2 \i \g \k - \i \p s - \ks \re (- \la)/T
        - \i \sum_{j=1}^{n_h} \th (\la - u_j) + \i \sum_{j=1}^{n_p} \th (\la - v_j) \\
	+ \int_{{\cal C}_{0, s}} \frac{\rd \m}{2 \p \i} \: K(\la - \m)
	  \ln ( 1 + \faq) (\m|\{u\}, \{v\}, \k)
\end{multline}
for the reciprocal $\faq = 1/\fa$ of the auxiliary function at
$u_k = x_k$ and $v_k = y_k$ and comparing with (\ref{resolventkernel}).
Then equation (\ref{straightnorm}) of the main text follows from
(\ref{detresolventform}) and (\ref{resaux}).

\Appendix{Details of the partial summation of the form-factor series
for the transverse two-point functions of the XXZ chain}
\label{app:sumxxz} \noindent
In this appendix we present a more detailed derivation of equation
(\ref{fftxxzfiniten}). Special attention will be payed to the
cancellation of the singularities in the integrand and to the issue of
spurious singularities related with possibly unphysical solutions of
the subsidiary conditions. In order to achieve a cancellation of
the singularities which is compatible with the usual formulation
of the multiple-residue theorem \cite{AiYu83,Range98} we shall explicitly
extract the poles of the functions $1 + \fa(\cdot|\{u\}, \{v\}, \k)$
and $1 + \faq(\cdot|\{u\}, \{v\}, \k)$. Concerning the issue of
spurious singularities, so far we can only state a set of assumptions
that would exclude them.

In the following the interior of the contours ${\cal C}_{0,s}$,
$\cbar_{0,s}$ will be denoted $\Om = \Int {\cal C}_{0, s}$,
$\Ombar = \Int \cbar_{0, s}$. It is convenient to combine the
sets $\{u\} = \{u_j\}_{j=1}^{n_h}$ and $\{v\} = \{v_j\}_{k=1}^{n_p}$
with $u_j \in \Om$, $v_k \in \Ombar$ and the twist parameter $\k'$
into triples ${\cal M} = (\{u\},\{v\},\k')$ and to write
$\fa(\cdot|\{u\}, \{v\}, \k') = \fa (\cdot|{\cal M})$ for short.
Similarly, solutions $\{x\}$, $\{y\}$ of the subsidiary conditions
(\ref{subsi}) and $\k'$ will be combined into ${\cal Z} =
(\{x\},\{y\},\k')$ such that $\fa(\cdot|\{x\}, \{y\}, \k') =
\fa (\cdot|{\cal Z})$. We also introduce the notation
\begin{equation}
     \chi (\la|{\cal M}) = 
	- \int_{{\cal C}_{0, s}} \frac{\rd \m}{2 \p \i} \: K(\la - \m)
	  \ln (1 + \fa) (\m|{\cal M})
\end{equation}
and restrict ourselves to $s = 1$ in the following.

Thanks to (\ref{nlieuv}),  the auxiliary function is presented as a ratio
\begin{equation} \label{factorfa}
     \fa(\la|{\cal M}) = \frac{\re^{\chi (\la|{\cal M})} g(\la|{\cal M})}
                              {h(\la|{\cal M})} \epc
\end{equation}
where the two functions
\begin{subequations}
\begin{align}
     & g(\la|{\cal M}) = (-1)^s q^{- 2 \k'} \de(\la)
        \biggl[\prod_{u \in \{u\}} \sh (\la - u + \i \g) \biggr]
	\biggl[\prod_{v \in \{v\}} \sh^2 (v - \la + \i \g) \biggr] \epc \\
     & h(\la|{\cal M}) = \a(\la)
        \biggl[\prod_{u \in \{u\}} \sh (\la - u + \i \g) \biggr]
	\biggl[\prod_{v \in \{v\}} \sh (v - \la + \i \g)
	                           \sh (\la - v + \i \g) \biggr]
\end{align}
\end{subequations}
are holomorphic in $\mathbb{C}$. Above, 
\begin{subequations}
\begin{align}
     \a (\la) & = \prod_{k=1}^{N+1} \sh(\la - \n_{2k}) \sh(\la - \n_{2k-1} - \i \g) \epc \\
     \de (\la) & = \prod_{k=1}^{N+1} \sh(\la - \n_{2k-1}) \sh(\la - \n_{2k} - \i \g) \epc
\end{align}
\end{subequations}
and $\n_k$, $k = 1, \dots, 2N + 2$, are defined according to (\ref{trotterdecomp}).

After elementary manipulations based on (\ref{factorfa}) the
amplitude $A_n^{-+} (\x|\k, \k')$ in (\ref{amp}) can be expressed as
\begin{equation} \label{apmkarol}
     A_n^{-+} (\x|\k, \k') = (-1)^{n_h}
        \frac{ \bigl[ \prod_{x \in \{x^{(n)}\}} \de (x) \bigr]
                         \bigl[ \prod_{y \in \{y^{(n)}\}} \a (y) \bigr]
			 }
             {\det
	\begin{vmatrix}
	   \6_{u_k} {\cal Y} (u_j|{\cal M}) & \6_{v_k} {\cal Y} (u_j|{\cal M}) \\
	   \6_{u_k} {\cal Y} (v_j|{\cal M}) & \6_{v_k} {\cal Y} (v_j|{\cal M}) 
	\end{vmatrix}_{{\cal M} = {\cal Z}_n}} \; {\cal F}^{-+} ({\cal Z}_n) \epp
\end{equation}
Here ${\cal Z}_n = (\{x^{(n)}\},\{y^{(n)}\},\k')$ is the triple associated
with the solution $\{x^{(n)}\},\{y^{(n)}\}$ to the subsidiary condition
(\ref{subsi}). Furthermore, we have introduced  
\begin{equation}
     {\cal Y} (\la|{\cal M}) = \re^{\chi (\la|{\cal M})} g(\la|{\cal M}) + h(\la|{\cal M})
\end{equation}
and the function ${\cal F}^{-+} ({\cal Z}_n)$ is of the form
\begin{equation}
     {\cal F}^{-+} ({\cal Z}_n) =
        \bigl({\cal H} \cdot {\cal D} \cdot {\cal W}
	       \cdot {\cal E}_1 \cdot {\cal E}_2 \bigr) ({\cal Z}_n)
\end{equation}
with
\begin{multline}
     {\cal H} ({\cal M}) =
        \frac{\overline{G}_+^- (\x) \overline{G}_-^+ (\x)}
	     {(q^{1 + \k' - \k} - q^{- 1 - \k' + \k})
              (q^{\k' - \k} - q^{- \k' + \k})} \\[1ex] \times \mspace{-3.mu}
        \frac{\det_{\rd m_+, {\cal C}_n} \bigl\{ 1 - \widehat{K}_{1-\k'+\k} \bigr\}
	      \det_{\rd m_-, {\cal C}_n} \bigl\{ 1 - \widehat{K}_{1+\k'-\k} \bigr\}}
	     {\det_{\rd m_0, {\cal C}_{0,1}} \bigl\{ 1 - \widehat{K} \bigr\}
              \det_{\rd m, {\cal C}_{0,1}} \bigl\{ 1 - \widehat{K} \bigr\}}
        \prod_{x \in \{u\}} (-1)^s q^{- 2\k'} \re^{\chi(x|{\cal M})}
\end{multline}
and
\begin{subequations}
\begin{align}
     & {\cal D} ({\cal M}) =
        \frac{\bigl[\prod_{u \ne u' \in \{u\}} \sh(u - u')\bigr]
	      \bigl[\prod_{v \ne v' \in \{v\}} \sh(v - v')\bigr]}
             {\prod_{u \in \{u\}} \prod_{v \in \{v\}}
	      \sh(u - v) \sh(v - u)} \epc \\[1ex]
     & {\cal W} ({\cal M}) = (-1)^{n_p}
        \frac{\prod_{u \in \{u\}} \prod_{v \in \{v\}}
	      \sh^2 (u - v + \i \g) \sh(v - u + \i \g)}
             {\prod_{u, u' \in \{u\}} \sh(u - u' + \i \g)} \epp
\end{align}
\end{subequations}
The functions $\overline{G}_{\pm}^{\mp}$ and the Fredholm determinants
are defined similar to equations \eqref{amp}-\eqref{kernelK} and
\eqref{measures1}-\eqref{measures2} of the main text: one only has to 
replace  $\fa (\cdot|{\cal Z})$ by $\fa (\cdot|{\cal M})$
and $\rho_n(\cdot|\kappa,\kappa')$ by $\rho(\cdot|{\cal M})$.
Moreover, writing
$\varnothing = (\emptyset, \emptyset, \k)$ and
\begin{equation}
     z (\la|{\cal M}) =
        \frac{\ln (1 + \fa) (\la|\varnothing)
	      - \ln (1 + \fa) (\la|{\cal M})}{2 \pi \i}
\end{equation}
we define
\begin{align}
     & {\cal E}_1 ({\cal M}) = (-1)^s q^{\k - \k'}
        \exp \biggl\{- \int_{{\cal C}_{0, 1}} \rd \la \int_{{\cal C}_{0, 1}'} \rd \m \:
	             \re'(\la - \m) z(\la|{\cal M}) z(\m|{\cal M}) \biggr\}
        \notag \\ & \mspace{72.mu} \times
	\prod_{\m \in \{v\}}
        \exp \biggl\{\int_{{\cal C}_{0, 1}} \rd \la \:
	             \bigl(\re(\la - \m) - \re(\m - \la)\bigr) z(\la|{\cal M}) \biggr\}
        \notag \\ & \mspace{72.mu} \times
	\prod_{\m \in \{u\}}
        \exp \biggl\{- \int_{{\cal C}_{0, 1}} \rd \la \:
	             \bigl(\re(\la - \m) - \re(\m - \la)\bigr) z(\la|{\cal M}) \biggr\}
		     \epc
\end{align}
where ${\cal C}_{0,1}' \subset {\cal C}_{0,1}$ infinitesimally, and
\begin{multline}
     {\cal E}_2 ({\cal M}) = \prod_{w \in \{u\} \cup \{v\}} \biggl[
        \Bigl[ \prod_{u \in \{u\}} \sh(w - u + \i \g) \Bigr]
        \Bigl[ \prod_{v \in \{v\}} \sh(v - w + \i \g) \Bigr] \\
        - (-1)^s q^{2(\k - \k')} \re^{\chi(w|\varnothing) - \chi(w|{\cal M})}
        \Bigl[ \prod_{u \in \{u\}} \sh(u - w + \i \g) \Bigr]
        \Bigl[ \prod_{v \in \{v\}} \sh(w - v + \i \g) \Bigr] \biggr].
\end{multline}

Inserting (\ref{apmkarol}) into our general expression (\ref{ffseriesnfinite})
for the form factor series of the two-point functions we obtain
\begin{multline} \label{ffpmf}
     \<\s_1^- \s_{m+1}^+ (t)\>_T =
        \lim_{\substack{N \rightarrow \infty\\ \e \rightarrow 0}}
        \sum_{\ell=1}^\infty
        \sum_{\substack{{\cal Z}_n : |\{x^{(n)}\}| \\  = |\{y^{(n)}\}| + 1 = \ell}}
        \frac{(-1)^{\ell} \re^{- \i h t} {\cal F}^{-+} ({\cal M})}
             {\det
	\begin{vmatrix}
	   \6_{u_k} {\cal Y} (u_j|{\cal M}) & \6_{v_k} {\cal Y} (u_j|{\cal M}) \\
	   \6_{u_k} {\cal Y} (v_j|{\cal M}) & \6_{v_k} {\cal Y} (v_j|{\cal M}) 
	\end{vmatrix}}
	\\[1ex] \times
	\Bigl[ \prod_{w \in \{u\}} \de(w) \Bigr]
	\Bigl[ \prod_{w \in \{v\}} \a(w) \Bigr] \r^m (0|{\cal M})
	\r^\frac{N}{2} (t_R/N|{\cal M}) \r^{- \frac{N}{2}} (- t_R/N|{\cal M})
	\biggr|_{{\cal M} = {\cal Z}_n} \epp
\end{multline}
Above, the sum runs, for fixed $\ell$, over all solutions $\{x^{(n)}\}$,
$\{y^{(n)}\}$ to the subsidiary conditions such that $|\{x^{(n)}\}| =
|\{y^{(n)}\}| + 1 = \ell$. The Jacobian put aside, the remaining functions
in the summand, viz.\ the factors
\begin{equation}
        {\cal F}^{-+} ({\cal M})
	\Bigl[ \prod_{z \in \{u\}} \de(z) \Bigr]
	\Bigl[ \prod_{z \in \{v\}} \a(z) \Bigr] \r^m (0|{\cal M})
	\r^\frac{N}{2} (t_R/N|{\cal M}) \r^{- \frac{N}{2}} (- t_R/N|{\cal M})
	\epc
\end{equation}
are already supposed to be analytic in $(\uv,\vv)$ belonging to the
natural domains $\Om^{n_h}\times \Ombar^{n_p}$ for the hole-type
variables $\uv = (u_1, \dots, u_{n_h})$ and for the particle-type
variables $\vv = (v_1, \dots, v_{n_p})$, where $n_p = n_h - 1$.
The explicit poles of $\r (t_R/N|{\cal M}) \r^{- 1} (- t_R/N|{\cal M})$ exist 
at $u_j = - t_R/N\,(1\le j \le n_h)$ and at $v_k = \i \g - t_R/N\,(1\le k \le n_p)$. 
See (\ref{evratuv}).
They  are canceled by the zeros of $\alpha(v_k)$ or $\delta(u_j)$. 
The linear equation (\ref{liegbar}) tells that explicitly  poles present in $\overline{G}_-^+ (0)$, resp.\
$\overline{G}_+^- (0)$ are located  at $u_j = 0$, resp.\ $v_k = \i \g$.
They are also canceled by the zeros of
$\r^m (0|{\cal M})$, provided that $m > 0$. In principle,
${\cal F}^{-+} ({\cal M})$ could also contain some singularities
stemming from the Fredholm determinants occurring in
${\cal H} ({\cal M}) $. However, we do not expect such kind of
complication and simply assume that it does not occur.

The summation in (\ref{ffpmf}) is equivalent to summing up all
solutions of the equation ${\cal Y} (z |{\cal M}) = 0$ with
$z \in \{u\} \cup \{v\}$, provided that the two summands
$\re^{\chi (\la|{\cal M})} g(\la|{\cal M})$ and $h(\la|{\cal M})$
do not vanish simultaneously on a solution. In principle, such
a situation might occur, e.g.\ due to the presence of the
common factor $\prod_{v \in \{v\}} \sh(v - \la + \i \g)$
in $h$ and $g$. We will make however the assumption that, even
if existing, such solutions do not contribute to the form factor
series, for instance because these also correspond to zeros
of ${\cal F}^{-+} ({\cal M})$ which are not manifestly appearing
in the formula or simply because these do not generate a
multi-dimensional residue. Based on this assumption we can use
the multi-dimensional residue formula \cite{AiYu83,Range98} to
recast the sum into the form
\begin{multline} \label{ffpmfint}
     \<\s_1^- \s_{m+1}^+ (t)\>_T =
        \lim_{\substack{N \rightarrow \infty\\ \e \rightarrow 0}}
        \sum_{\ell=1}^\infty
        \frac{(-1)^\ell \re^{- \i h t}}{\ell! (\ell-1)!}
	\int_{{\cal S}_{\cal Y}^{(\ell, \e)}}
	\frac{\rd^\ell u \, \rd^{\ell-1} v}{(2 \p \i)^{2\ell - 1}}
        \frac{{\cal F}^{-+} ({\cal M})}
             {\prod_{w \in \{u\} \cup \{v\}} {\cal Y} (w|{\cal M})}
	\\[1ex] \times
	\Bigl[ \prod_{w \in \{u\}} \de(w) \Bigr] \Bigl[ \prod_{w \in \{v\}} \a(w) \Bigr]
	\r^m (0|{\cal M})
	\r^\frac{N}{2} (t_R/N|{\cal M}) \r^{- \frac{N}{2}} (- t_R/N|{\cal M}) \epp
\end{multline}
Here $\e > 0$, and ${\cal S}_{\cal Y}^{(\ell, \e)}$ is the skeleton
of ${\cal Y}$ defined as
\begin{equation}\label{eq:def_skeleton}
     {\cal S}_{\cal Y}^{(\ell, \e)} =
        \bigl\{ (\uv, \vv) \in \Om^\ell \times \Ombar^{\ell - 1} \big|
	        |{\cal Y} (u_j|{\cal M})| = |{\cal Y} (v_k|{\cal M})| = \e \bigr\} \epp
\end{equation}

We assume that this skeleton can be deformed
into ${\cal C}_{0,1}^\ell \times \cbar_{0,1}^{\ell - 1}$.
Then the series representation (\ref{fftxxzfiniten}) in the
main text easily follows.

\Appendix{Details of the derivation of the form-factor expansions
for the XX chain} \label{app:detailsxx} \noindent
For our derivation of form-factor formulae for the XX chain that
are suitable for taking the Trotter limit we recall Slavnov's scalar
product formula for the XXZ model \cite{Slavnov89}. For any set
of Bethe roots $\{\la_j^{(n)}\}_{j=1}^M$ and its associated auxiliary
function $\fa_n$ it takes the form
\begin{multline} \label{slavnov}
     \<\Ps_n|B(\m_M) \dots B(\m_1)|0\> = \<0|C(\m_1) \dots C(\m_M) |\Ps_n\> \\[1ex]
        = \frac{\prod_{j=1}^M d(\la_j^{(n)}) a(\m_j)
                \prod_{k=1}^M \sh(\la_j^{(n)} - \m_k - \i \g)}
               { \prod_{1 \le j < k \le M} \sh(\la_j^{(n)} - \la_k^{(n)}) \sh(\m_k - \m_j)}
	  \\ \times
          \det_M \bigl\{ K(\la_j^{(n)} - \m_k) - \re(\m_k - \la_j^{(n)})
	                (1 + \fa_n (\m_k|\k)) \bigr\}
	  \epp
\end{multline}
The set $\{\m_j\}_{j=1}^M$ is still arbitrary in this formula.
Sending $\m_j \rightarrow \la_j^{(n)}$ we get the `norm formula' for
the eigenstates (\ref{evs}).
\subsection{Longitudinal case}
Equation (\ref{slavnov}) implies a formula for the ratio of the scalar
product divided by the `square of the norm' which in the XX limit
simplifies due to (\ref{ekxx}),
\begin{multline} \label{slavnovxx}
     \frac{\<\Ps_n| B(\m_M) \dots B(\m_1)|0\>}{\<\Ps_n|\Ps_n\>}
        = \biggl[ \prod_{j=1}^M \frac{a(\m_j)}{a(\la_j^{(n)}) \fa_n' (\la_j^{(n)}|\k)}
                  \prod_{k=1}^M \frac{\ch(\la_j^{(n)} - \m_k)}
		                     {\ch(\la_j^{(n)} - \la_k^{(n)})} \biggr] \\[1ex] \times
          \biggl[ \prod_{1 \le j < k \le M}
	          \frac{\sh(\la_j^{(n)} - \la_k^{(n)})} {\sh(\m_j - \m_k)} \biggr]
          \times \det_M \biggl\{ \frac{2 \bigl(1 + \fa_n (\m_k|\k)\bigr)}
	                              {\sh(2(\m_k - \la_j^{(n)}))} \biggr\} \epp
\end{multline}

If now $B(\m_M) \dots B(\m_1) |0\>$ is a Bethe vector as well, it
has the same pseudo-spin $s$ as $\<\Ps_n|$, since it has the same number
of Bethe roots. It follows that
\begin{equation}
     \frac{2 \bigl(1 + \fa_n (\m_k|\k)\bigr)}{\sh(2(\m_k - \la_j^{(n)}))} =
        \lim_{\m \rightarrow \m_k}
	\frac{2 \bigl(1 + \fa_n (\m|\k)\bigr)}{\sh(2(\m - \la_j^{(n)}))} = 0
\end{equation}
unless $\m_k = \la_j^{(n)}$ for some $j$, in which case
\begin{equation}
     \lim_{\m \rightarrow \m_k}
	\frac{2 \bigl(1 + \fa_n (\m|\k)\bigr)}{\sh(2(\m - \la_j^{(n)}))}
	= \fa_n' (\la_j^{(n)}|\k) \epp
\end{equation}
Thus, if $\m_k \notin \{\la_j^{(n)}\}_{j=1}^M$, then a row in
the determinant on the right hand side of equation (\ref{slavnovxx})
vanishes, and the normalized scalar product is equal to zero. It
follows that $\<\Ps_n|\Ps_m\> = 0$ for any two different sets of
Bethe roots $\{\la_j^{(n)}\}_{j=1}^M$ and $\{\la_j^{(m)}\}_{j=1}^M$
(which proves completeness since the number of solutions equals
the dimension of the Hilbert space and $\<\Ps_n|\Ps_n\> \ne 0$).

Using this fact as well as the Yang-Baxter algebra relations
(\ref{yba}) we conclude that
\begin{multline} \label{aaction}
     \frac{\<\Ps_n|A(\x)|\Ps_m\>}{\<\Ps_n|\Ps_n\>}
       = \sum_{j=1}^M a(\la_j^{(m)}) \frac{c(\x,\la_j^{(m)})}{b(\x,\la_j^{(m)})} \\ \times
         \biggl[
	   \prod_{\substack{k = 1\\k \ne j}}^M \frac{1}{b(\la_k^{(m)},\la_j^{(m)})} \biggr]
     \frac{\<\Ps_n| B(\x) \bigl[
           \prod_{k = 1, k \ne j}^M B(\la_k^{(m)}) \bigr]|0\>}
          {\<\Ps_n|\Ps_n\>}
\end{multline}
if $\{\la_j^{(n)}\}_{j=1}^M$ and $\{\la_j^{(m)}\}_{j=1}^M$ are
two different solutions of the Bethe Ansatz equations
$1 + \fa_n (\la|\k) = 0$. Following the same reasoning as above,
each term on the right hand side of (\ref{aaction}) can only be
non-zero if $\{\la_k^{(m)}\}_{k=1, k \ne j}^M \subset
\{\la_j^{(n)}\}_{j=1}^M$. Since the two solutions of the Bethe
Ansatz equations are different, at least one $\la_\ell^{(m)} \notin
\{\la_j^{(n)}\}_{j=1}^M$, implying that only the summand with $j = \ell$
can be non-zero. Without any loss of generality we then assume that
$\la_j^{(m)} = \la_j^{(n)}$ for $j = 1, \dots, M - 1$. For such states
(\ref{slavnovxx}) and (\ref{aaction}) imply that
\begin{multline}
     \frac{\<\Ps_n|A(\x)|\Ps_m\>}{\<\Ps_n|\Ps_n\>}
       = a(\la_M^{(m)}) \frac{c(\x,\la_M^{(m)})}{b(\x,\la_M^{(m)})}
         \biggl[ \prod_{j = 1}^{M-1} \frac{1}{b(\la_j^{(n)},\la_M^{(m)})} \biggr]
	 \frac{a(\x)}{a(\la_M^{(n)}) \fa_n' (\la_M^{(n)}|\k)}
		 \\ \times
	 \biggl[ \prod_{j=1}^M \frac{\ch(\la_j^{(n)} - \x)}{\ch(\la_j^{(n)} - \la_M^{(n)})}
	         \biggr]
         \biggl[ \prod_{k = 1}^{M-1}
	         \frac{\sh(\la_j^{(n)} - \la_M^{(n)})}{\sh(\la_j^{(n)} - \x)}
	         \biggr] \re(\x - \la_M^{(n)}) \bigl( 1 + \fa_n (\x|\k) \bigr) \epp
\end{multline}

In order to calculate the amplitudes $A_n$ we consider two cases. First,
$\{\la_j^{(n)}\}_{j=1}^M$ are the Bethe roots of the dominant state. Then
$s = 0$ and $\la^p := \la_M^{(m)}$ has imaginary part larger than $\p/4$, hence
is a particle, while $\la^h := \la_M^{(n)}$ is missing in the set
$\{\la_j^{(m)}\}_{j=1}^M$ and is a hole. It follows that
\begin{multline} \label{ampzeron}
     \frac{\<\Ps_0|A(\x)|\Ps_n\>}{\<\Ps_0|\Ps_0\>}
       = a(\x) \bigl( 1 + \fa_0 (\x|\k) \bigr)
         \frac{\re(\x - \la^h)}{\fa_0' (\la^h|\k)}
	 \frac{a(\la^p)}{a(\la^h)} \\ \times
	 \frac{c(\x,\la^p)}{b(\x,\la^p)} \ch(\la^h - \x)
         \biggl[ \prod_{j = 1}^{M-1}
	         \frac{- \i \ch(\la_j^{(n)} - \la^p)}{\sh(\la_j^{(n)} - \la^p)}
	         \frac{\ch(\la_j^{(n)} - \x)}{\ch(\la_j^{(n)} - \la^h)}
                 \frac{\sh(\la_j^{(n)} - \la^h)}{\sh(\la_j^{(n)} - \x)} \biggr] \epp
\end{multline}
In the second case we take $\{\la_j^{(m)}\}_{j=1}^M$ as the Bethe
roots of the dominant state. Then $\{\la_j^{(n)}\}_{j=1}^M$ has
to describe an excited state with one particle $\la^p = \la_M^{(n)}$
and one hole $\la^h = \la_M^{(m)}$. Thus,
\begin{multline} \label{ampnzero}
     \frac{\<\Ps_n|A(\x)|\Ps_0\>}{\<\Ps_n|\Ps_n\>}
       = a(\x) \bigl( 1 + \fa_0 (\x|\k) \bigr)
         \frac{\re(\x - \la^p)}{\fa_0' (\la^p|\k)}
	 \frac{a(\la^h)}{a(\la^p)} \\ \times
	 \frac{c(\x,\la^h)}{b(\x,\la^h)} \ch(\la^p - \x)
         \biggl[ \prod_{j = 1}^{M-1}
	         \frac{- \i \ch(\la_j^{(n)} - \la^h)}{\sh(\la_j^{(n)} - \la^h)}
	         \frac{\ch(\la_j^{(n)} - \x)}{\ch(\la_j^{(n)} - \la^p)}
                 \frac{\sh(\la_j^{(n)} - \la^p)}{\sh(\la_j^{(n)} - \x)} \biggr] \epp
\end{multline}

Multiplying (\ref{ampzeron}) and (\ref{ampnzero}) and using the
formula (\ref{evasxxz}) for the eigenvalues we arrive at
\begin{equation}
     \frac{\<\Ps_0|A(\x)|\Ps_n\>}{\<\Ps_0|\Ps_0\> \La_n (\x|\k)}
     \frac{\<\Ps_n|A(\x)|\Ps_0\>}{\<\Ps_n|\Ps_n\> \La_0 (\x|\k)} =
        \frac{\re(\x - \la^h)}{\fa_0' (\la^h|\k)}
        \frac{\re(\x - \la^p)}{\fa_0' (\la^p|\k)} \epc
\end{equation}
which is our final expression for the amplitudes at finite Trotter
number. The Trotter limit and the limit $\e \rightarrow 0$ are determined
by equation (\ref{atrotterxx}).

\subsection{Transverse case}
In the transverse case we have to evaluate the amplitudes $A_n^{-+} (\x)$
defined in equation (\ref{defatrans}) of the main text. Let us suppress
superscripts for short in this section and denote the Bethe roots of
the dominant state simply by $\{\la_j\}_{j=1}^M$, $M = N + 1$. The only
excited states $|\Ps_n\>$ that lead to non-zero amplitudes are those
with $M - 1$ Bethe roots. In this section we denote them by
$\{\m_k\}_{k=1}^{M-1}$. They correspond to spin $s = 1$ (see
(\ref{pseudospin})) and are zeros of $1 + \fa_n$, where $\fa_n =
- \fa_0$. Using the Slavnov formula (\ref{slavnov}) and the equation
(\ref{evasxxz}) for the eigenvalues and setting $\g = \p/2$ we
find that
\begin{multline} \label{apm1}
     A_n^{-+} (\x) = (-1)^{M-1} \frac{1 + \fa_0 (\x|\k)}{1 + \fa_n (\x|\k)}
        \biggl[ \prod_{j=1}^M \frac{2}{\fa_0' (\la_j|\k)} \biggr]
        \biggl[ \prod_{k=1}^{M-1} \frac{2}{\fa_n' (\m_k|\k)} \biggr] \\ \times
	\frac{\prod_{k=1}^{M-1} \sh\bigl(2(\x - \m_k)\bigr)}
	     {\prod_{j=1}^M \sh\bigl(2(\x - \la_j)\bigr)} \:
	     {\cal D} \bigl( \{\la_j\}_{j=1}^M, \{\m_k\}_{k=1}^{M-1} \bigr) \epc
\end{multline}
where for any two mutually distinct sets of complex numbers ${\cal D}$
is defined in (\ref{defd}).

We would like to rewrite (\ref{apm1}) in a form that allows us to take
the Trotter limit. Let us start with introducing some useful notation.
Let
\begin{equation}
     {\cal B}_s = \Bigl\{ \la \in {\mathbb C} \Big| \fa_0 (\la|\k) = (-1)^{s+1},\
                   |\Im \la| \le \p/4,\ \Re \la < 0\ \text{if}\
		   \Im \la = \p/4 \Bigr\} \epc
\end{equation}
where $s = 0, 1$. Then ${\cal B}_0 = \{\la_j\}_{j=1}^M$ is the set
of Bethe roots of the dominant state. We further define sets $\cal P$
of `particles' and $\cal H$ of `holes' by
\begin{equation}
     {\cal P} = \{\m_k\}_{k=1}^{M-1} \setminus
                  \bigl\{ {\cal B}_1 \cap \{\m_k\}_{k=1}^{M-1} \bigr\} \epc \qd
     {\cal H} = {\cal B}_1 \setminus
                  \bigl\{ {\cal B}_1 \cap \{\m_k\}_{k=1}^{M-1} \bigr\} \epp
\end{equation}
All pseudo-spin 1 excitations are uniquely classified by set of particles
and holes. Slightly abusing the notion of the difference of two sets
${\cal X}$ and ${\cal Y}$ we introduce the notation
\begin{equation}
     \prod_{\la \in {\cal X} \setminus {\cal Y}} f(\la) =
        \frac{\prod_{\la \in {\cal X}} f(\la)}
	     {\prod_{\la \in {\cal Y}} f(\la)}
\end{equation}
which will turn out to be convenient in the following calculations.
We shall also make use of the function
\begin{equation}
     {\bm 1}_{\rm condition} = \begin{cases}
                                  1 & \text{if condition is satisfied} \\
				  0 & \text{else.}
                               \end{cases}
\end{equation}

Using the above defined notation we separate the particles and holes
from the products in (\ref{apm1}),
\begin{multline} \label{apm2}
     A_n^{-+} (\x) = (-1)^{M-1} \frac{1 + \fa_0 (\x|\k)}{1 + \fa_n (\x|\k)}
        \biggl[ \prod_{\la \in {\cal B}_0} \frac{2}{\fa_0' (\la|\k)} \biggr]
        \biggl[ \prod_{\la \in {\cal B}_1} \frac{2}{\fa_n' (\la|\k)} \biggr]
        \biggl[ \prod_{\la \in {\cal P} \setminus {\cal H}} \frac{2}{\fa_n' (\la|\k)}
	        \biggr] \\ \times
	\biggl[ \prod_{\la \in {\cal B}_1 \setminus {\cal B}_0}
	           \sh\bigl(2(\x - \la)\bigr) \biggr]
	\biggl[ \prod_{\la \in {\cal P} \setminus {\cal H}}
	           \sh\bigl(2(\x - \la)\bigr) \biggr]
	\biggl[ \prod_{\la \in {\cal P} \setminus {\cal H}} \:
	        \prod_{\substack{\m \in {\cal B}_1 \setminus {\cal B}_0 \\ \m \ne \la}}
		       \sh^2 (\la - \m) \biggr] \\ \times
	     {\cal D} \bigl( {\cal B}_0, {\cal B}_1 \bigr)
	     {\cal D} \bigl( {\cal P}, {\cal H} \bigr) \epp
\end{multline}
Then logarithms of products over ${\cal B}_0$ and ${\cal B}_1$ can
be transformed into integrals involving the auxiliary functions $\fa_0$
and $\fa_n$. We choose a point $x_R$ on ${\cal C}$ in such a way
that $z(\la)$ defined in $(\ref{defz})$ is continuous on ${\cal C}$
with the possible exception of $\la = x_R$. It follows that
\begin{multline}
     \prod_{\la \in {\cal B}_1 \setminus {\cal B}_0} \sh\bigl(2(\x - \la)\bigr) =
        \sh^{M_1 - M} \bigl(2 (\x - x_R)\bigr)
	\biggl( \frac{1 + \fa_n (\x|\k)}{1 + \fa_0 (\x|\k)}
	        \biggr)^{{\bf 1}_{\x \in {\rm Int} ({\cal C})}} \\ \times
        \exp \biggl\{ - 2 \int_{\cal C} \rd \m \cth \bigl(2 (\x - \m)\bigr) z(\m) \biggr\}
	\epc
\end{multline}
where $M_1 = |{\cal B}_1|$ is the number of zeros of $1 + \fa_n$
inside ${\cal C}$ and $z$ is defined in equation~(\ref{defz}). By
${\rm Int} ({\cal C})$ we mean the interior of the contour $\cal C$.
With the remaining products over ${\cal B}_0$ and ${\cal B}_1$ we
proceed in a similar way. We have to be careful with the omissions
though. To treat them properly we introduce a small regularization
parameter $\de \in {\mathbb C}$. We have
\begin{multline}
     \prod_{\m \in {\cal B}_1 \setminus {\cal B}_0} \sh(\la - \m + \de) =
        \sh^{M_1 - M} (\la - x_R + \delta)
	\biggl( \frac{1 + \fa_n (\la + \de|\k)}{1 + \fa_0 (\la + \de|\k)}
	        \biggr)^{{\bf 1}_{(\la + \de) \in {\rm Int} ({\cal C})}} \\ \times
        \exp \biggl\{ - \int_{\cal C} \rd \m \cth (\la - \m + \de) z(\m) \biggr\}
\end{multline}
and therefore
\begin{multline}
     \prod_{\la \in {\cal P} \setminus {\cal H}} \:
     \prod_{\substack{\m \in {\cal B}_1 \setminus {\cal B}_0 \\ \m \ne \la}}
		       \sh^2 (\la - \m) =
     \lim_{\delta \rightarrow 0} \: \de^{2 |{\cal H}|}
     \prod_{\la \in {\cal P} \setminus {\cal H}} \:
     \prod_{\m \in {\cal B}_1 \setminus {\cal B}_0} \sh^2 (\la - \m + \de) \\
     = \lim_{\delta \rightarrow 0} \: \de^{2 |{\cal H}|}
       \prod_{\la \in {\cal P} \setminus {\cal H}}
        \sh^{2(M_1 - M)} (\la - x_R + \de)
	\biggl( \frac{1 + \fa_n (\la + \de|\k)}{1 + \fa_0 (\la + \de|\k)}
	        \biggr)^{2 \times {\bf 1}_{(\la + \de) \in {\rm Int} ({\cal C})}} \\ \times
        \exp \biggl\{ - 2 \int_{\cal C} \rd \m \cth (\la - \m + \de) z(\m) \biggr\} \\
     = \biggl[ \prod_{\la \in {\cal H}} \biggl( \frac{2}{\fa_n' (\la|\k)} \biggr)^2 \biggr]
       \prod_{\la \in {\cal P} \setminus {\cal H}}
        \sh^{2(M_1 - M)} (\la - x_R)
        \exp \biggl\{ - 2 \int_{\cal C} \rd \m \cth (\la - \m) z(\m) \biggr\} \epp
\end{multline}
Similarly
\begin{multline}
     {\cal D} \bigl( {\cal B}_0, {\cal B}_1 \bigr) =
        (-1)^{M_{01}} \lim_{\delta \rightarrow 0} \: \de^{- M - M_1}
        \prod_{\la \in {\cal B}_1 \setminus {\cal B}_0}
        \prod_{\m \in {\cal B}_1 \setminus {\cal B}_0} \sh(\la - \m + \de) \\
     = (-1)^{M_{01}} \lim_{\delta \rightarrow 0} \: \de^{- M - M_1}
        \prod_{\la \in {\cal B}_1 \setminus {\cal B}_0}
        \sh^{M_1 - M} (\la - x_R + \delta)
	\biggl( \frac{1 + \fa_n (\la + \de|\k)}{1 + \fa_0 (\la + \de|\k)}
	        \biggr)^{{\bf 1}_{(\la + \de) \in {\rm Int} ({\cal C})}} \\ \times
        \exp \biggl\{ - \int_{\cal C} \rd \m \cth (\la - \m + \de) z(\m) \biggr\} \\[1ex]
     = (-1)^{M_{01}}
       \biggl[ \prod_{\la \in {\cal B}_0} \frac{\fa_0' (\la|\k)}{2} \biggr]
       \biggl[ \prod_{\la \in {\cal B}_1} \frac{\fa_n' (\la|\k)}{2} \biggr]
       \biggl[ \prod_{\la \in {\cal B}_1 \setminus {\cal B}_0}
               \sh^{M_1 - M} (\la - x_R) \biggr] \\
       \exp \biggl\{ - \int_{{\cal C}' \subset {\cal C}} \rd \la
                       \int_{{\cal C}} \rd \m \cth' (\la - \m) z(\m) z(\la)
		     + (M_1 - M) \int_{{\cal C}} \rd \m \cth(\m - x_R) z(\m) \biggr\}
		     \epc
\end{multline}
where ${\cal C}'$ is inside ${\cal C}$ in such a way that it still
encloses all $\la \in {\cal B}_0$ and no other zeros of $1 + \fa_0$,
and where
\begin{equation}
     M_{01} = \2 (M - M_1)(M + M_1 -1) + M M_1 \epp
\end{equation}
At least for large enough Trotter number the choice of the
the contour ${\cal C}$ as depicted in figure~\ref{fig:xx_transverse_contour}
assures that $M_1 = M$. It follows that $M - |{\cal H}| + |{\cal P}|
= M - 1$ implying that $|{\cal H}| - |{\cal P}| = s = 1$.
When $M_1 = M$, many of the above expressions simplify. Inserting
the simplified expressions into (\ref{apm2}) we arrive at
equation (\ref{ampspmxx}) of the main text.

}

\bibliographystyle{amsplain}
\bibliography{hub}

\end{document}